\documentclass[11pt,epsfig]{JINST}
\newcommand{\be}{\begin{equation}}
\newcommand{\ee}{\end{equation}}

\def\p0f{\pi^0}

\def\K0barf{\overline{\mathrm{K^0}}}
\def\KL0f{\mathrm{K_L^0}}
\def\KS0f{\mathrm{K_S^0}}

\title{A hardware implementation of Region-of-Interest selection in LAr-TPC
 for data reduction and triggering}

\author{B. Baibussinov$^a$, S. Centro$^a$$\thanks{Corresponding author}$,
  K. Cieslik$^{a,b}$, D. Dequal$^a$, C. Farnese$^a$, A. Fava$^a$, 
  
   D. Gibin$^a$, A. Guglielmi$^a$, G. Meng$^a$, F. Pietropaolo$^a$, C. Rubbia$^{c,d}$, 
   
   E. Scantamburlo$^a$, F. Varanini$^a$, S. Ventura$^a$\\
   
\llap{$^a$} Istituto Nazionale di Fisica Nucleare and Dept. of Physics  ``G. Galilei",\\
  Via Marzolo 8 35131 Padova, Italy\\
\llap{$^b$} On leave form Institut Fizyki Jadrowej PAN,\\
  Krakov, Poland\\
 \llap{$^c$}  Laboratori Nazionali del Gran Sasso dell'INFN,\\
  I-67010 Assergi (AQ), Italy\\
\llap{$^d$} CERN, European Laboratory for Particle Physics\\
 CH-1211 Geneve 23, Switzerland\\
  E-mail: \email{sandro.centro@pd.infn.it}}

\abstract{Large Liquid Argon TPC detectors in the range of multikton mass for neutrino and astroparticle physics require 
the extraction and treatment of signals from some $10^5$ wires. In order to enlarge the throughtput of the DAQ system an 
on-line lossless data compression has been realized reducing almost a factor 4 the data flow. Moreover a trigger 
system based on a new efficient on-line identification algorithm of wire hits was studied, implemented on the actual ICARUS
digital read-out boards and fully tested on the ICARINO LAr-TPC facility operated at LNL INFN Laboratory  with cosmic-rays.
Capability to trigger isolated low energy events down to 1 MeV visible energy was also demonstrated.

\vskip 1cm
\begin{center}
\it{To be submitted to JINST}
\end{center}

}


\begin{document}

\newpage

\section{Introduction}

 Developments in neutrino and astroparticle physics will require the realization of gigantic detectors with mass 
 of order of hundreds kton or megaton to be installed in underground laboratories. Such an apparatus will be addressed 
 to study long base-line neutrino oscillations at accelerators as well as from cosmic sources including Supernovae, and to 
 test the barion matter stability searching for proton decay. Among the proposed detection techniques, the Liquid Argon
 TPC detector (LAr-TPC) is expected to play a special role, accounting for the excellent event imaging capability and the
 superior performance concerning event detection efficiency and identification which allows to keep the detector size in 
 the multikton mass range with respect to the megaton masses required for example by the water Cherenkov technique. Thus 
 the efforts linked to the underground site excavation and infrastructures, as well as the engineering needs, can be 
 largely reduced. The successful starting operations of ICARUS-T600 \cite{ICARUS-T600} 600 ton LAr-TPC at INFN LNGS
 underground laboratory using the CERN CNGS neutrino beam, represent a milestone towards the realization of large 
 LAr-TPC mass detectors for neutrino and astroparticle physics.

 The extrapolation to multikton mass LAr-TPC detector requires some improvements in different aspects of the technology,
 such as cryogenic plant, LAr purity and detector read-out \cite{MODULAr}. Moreover extraction and handling of the 
 corresponding huge amount of data will constitute a not negligible aspect to be addressed in the detector exploitation. 
 In the present ICARUS-T600 about 54000 channels of the wire TPC are read-out.
 To increase the bandwidth of the DAQ system two improvements have been studied for the TPC wire signal read-out: an on-line 
 lossless factor 4 data compression, and a new algorithm to detect the local ROIs (Region of Interest) of each event avoiding 
 the full acquisition of the detector. In particular the second one will allow triggering isolated low energy events down to 1 
 MeV visible energy using directly the wire signals.

 In the following the algorithm description and implementation on the  TPC read-out board is described as well as the performance, 
 tested with the ICARINO LAr-TPC facility in the framework of the ICARUS R\&D activity at the INFN-LNL laboratories in Legnaro (Italy).

\section{Cosmic-ray test-run with ICARINO test-facility at LNL}
 
\begin{figure}[!htb]
\begin{center}
\includegraphics[width=0.4\textwidth]{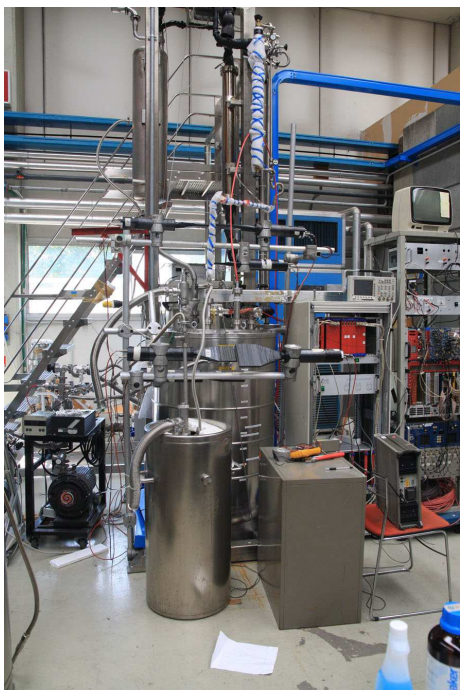}
 \caption{The ICARINO test-facility set-up in LNL.}
\label{icarino_1}
\end{center}
\end{figure}

\begin{figure}[!htb]
\begin{center}
\includegraphics[width=0.4\textwidth]{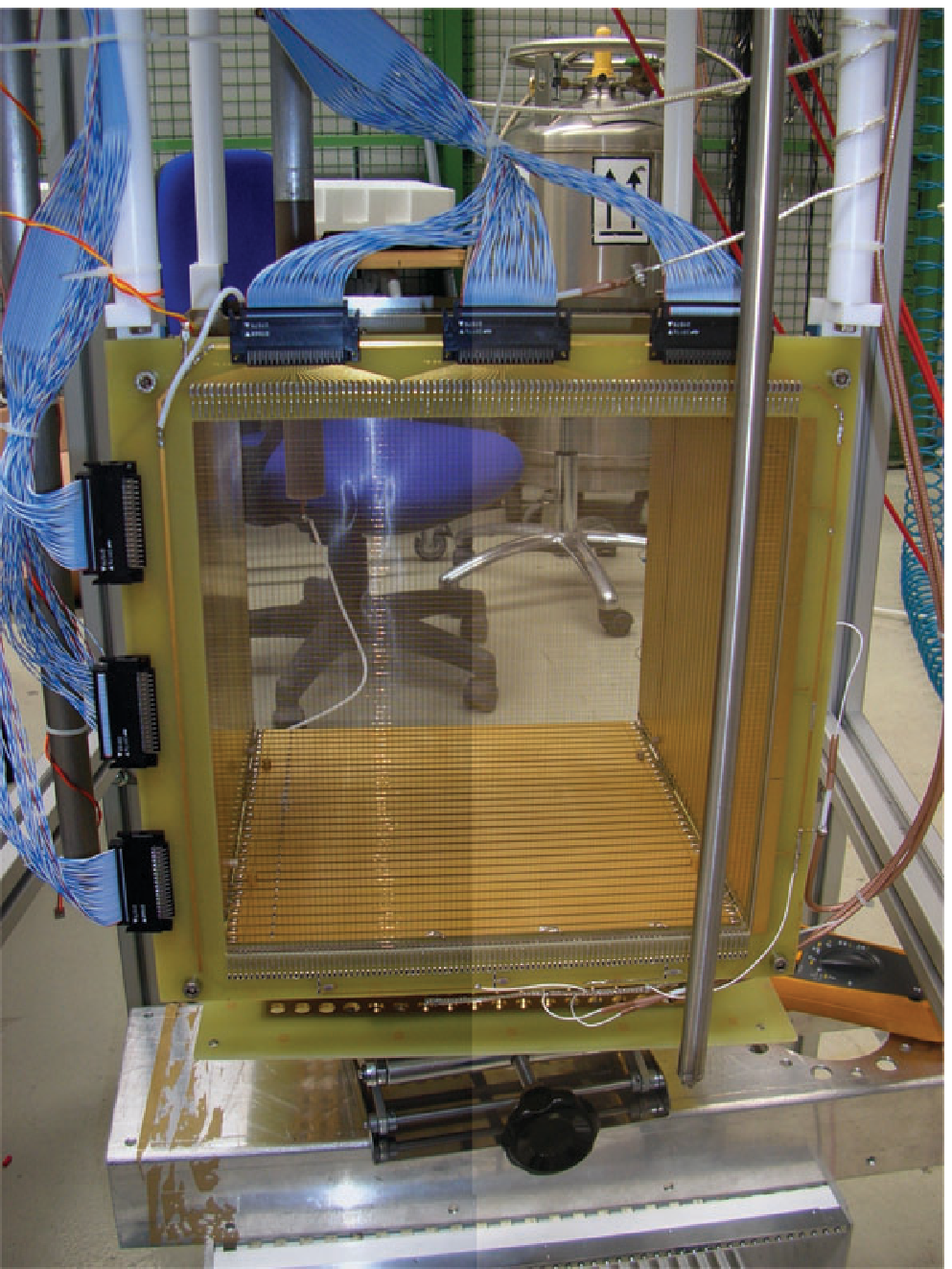}
\includegraphics[width=0.4\textwidth]{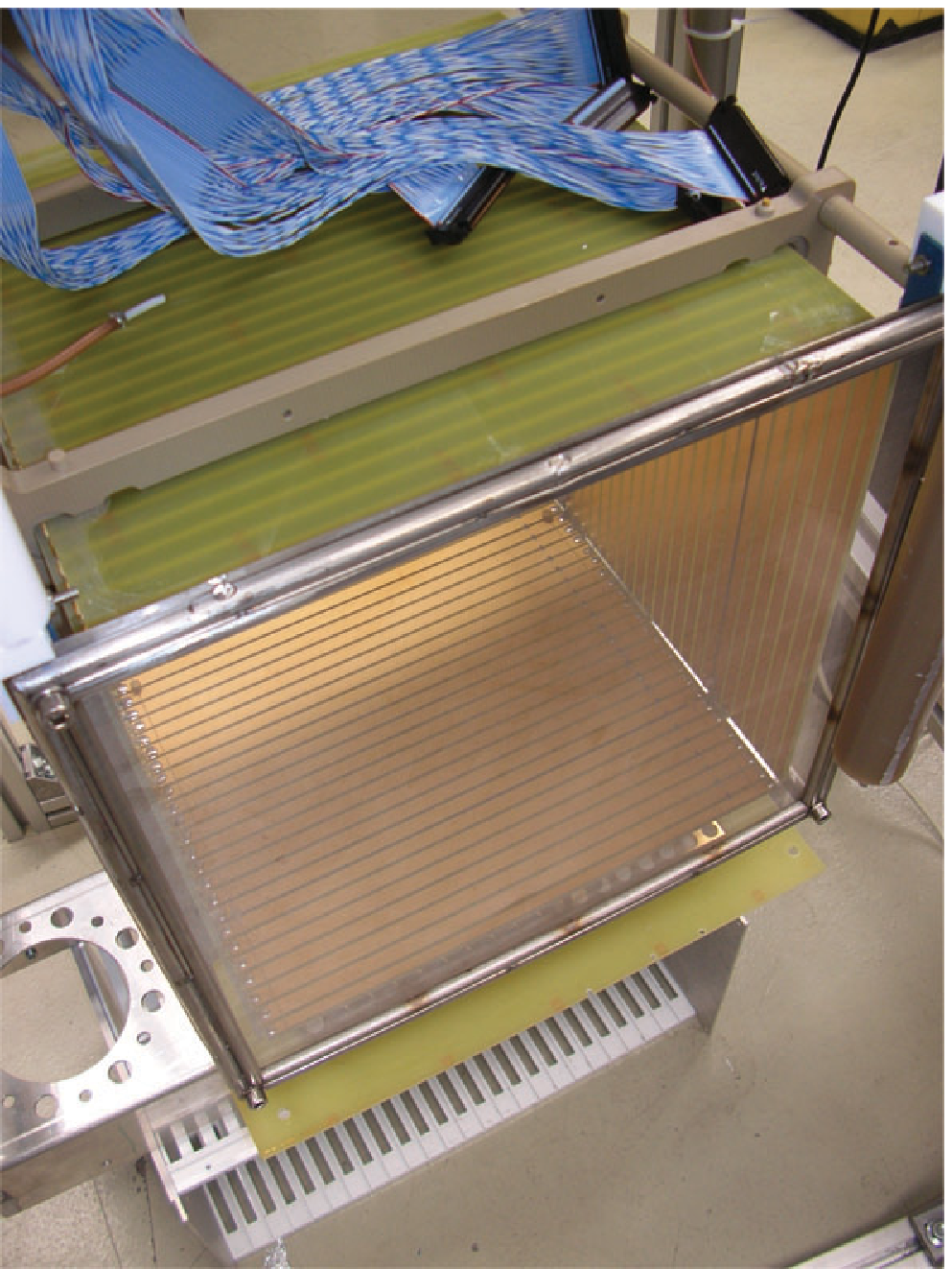}
 \caption{Images of LAr-TPC at LNL: anode (left) and cathode (right) structure of the TPC.}
\label{icarino_2}
\end{center}
\end{figure}

 A detailed description of the ICARINO experimental set-up (Fig. \ref{icarino_1}) can be found elsewhere \cite{icarino}. The TPC
 consists of two vertical electrode planes, $32.6 \times 32.6$ cm$^2$, acting as anode and cathode, laterally delimited by 4 
 vetronite $29.4 \times 29.4$ cm$^2$ boards supporting the field-shaping electrodes (30 strips of gold-plated copper) resulting 
 in a 38 kg LAr active mass (Fig. \ref{icarino_2}). In order to facilitate the LAr re-circulation in the TPC active volume, 
 the cathode at the front of the chamber has been realized with a thin etched stainless steel grid.

The chamber is contained in a stainless steel cylindrical vessel, whose upper face is an ultra high vacuum flange hosting 
the feed-throughs for vacuum, LAr filling and re-circulation, high voltage cables and read-out electronics. The whole 
detector vessel is contained in an open-air stainless steel dewar, which is initially filled with commercial LAr acting 
as cryogenic bath for the ultrapure LAr injected in the detector vessel.
   
The read-out anode's electrodes facing the drift volume are two parallel 96 stainless steel wire planes spaced by 3 mm 
and with 3 mm pitch: the first one works in induction mode and is made of vertical wires while the second one collects 
the drifting electrons and is made of horizontal wires. A third wire plane electrically biased, called grid, is inserted 
3.5 mm in front of the Induction wires. It acts as an electromagnetic shield  improving the induction signal sharpness
and reducing the noise due to the HV biasing. Finally, another thin stainless steel grid has been inserted behind the Collection
wire plane and put to ground to confine the electric field around the wires.
  
A -14.8 kV voltage is applied on the cathode corresponding to a uniform electric field of 474 V/cm in the drift volume.
In order to ensure a full transparency of the Grid and of the Induction plane to drifting electrons, the potential is fixed at 350 V 
for the Collection wires, - 100 V for the Induction wires and - 350 V for the Grid. This set of values was chosen after
 a careful study on the reconstructed muon tracks collected in several preliminary test runs, optimizing the electric 
 field uniformity in the whole drift region.\\
 
The TPC vessel is filled with ultra pure LAr, following the procedure described in \cite{icarino}. First the chamber is
 evacuated to at least $10^{-3}$ Pa with turbo-molecular pumping system for several days, to maximize degassing of the 
 detector materials. Then the chamber is cooled down in the LAr bath as fast as possible and is filled with commercial 
 LAr purified by the Oxysorb/Hydrosorb filter. An additional small evacuated tank immersed into a LAr bath, acting as a 
 cryogenic trap, is connected to the TPC vessel during the first phase of the filling to remove the gaseous Argon, 
 produced by the evaporation on the warm inner detector surfaces and possibly contaminated by degassing.\\
 This additional vessel is disconnected soon after the completion of the filling procedure. 
 
 The level of the LAr cryogenic bath, in which the detector vessel is immersed, is lowered with respect of that in the inner detector, to increase the heat losses and allow the recirculation system to start.
 The gaseous Argon (GAr), evaporating from the liquid in the detector vessel, is purified by the Oxysorb/Hydrosorb filter,
 condensed in a passive heat exchanger (LAr bath) and finally re-injected at the bottom of the detector. The total LAr volume 
 is completely recirculated every 2.5 days.

\begin{figure}[htbp]
\begin{center}
\includegraphics[width=12cm,angle=-90]{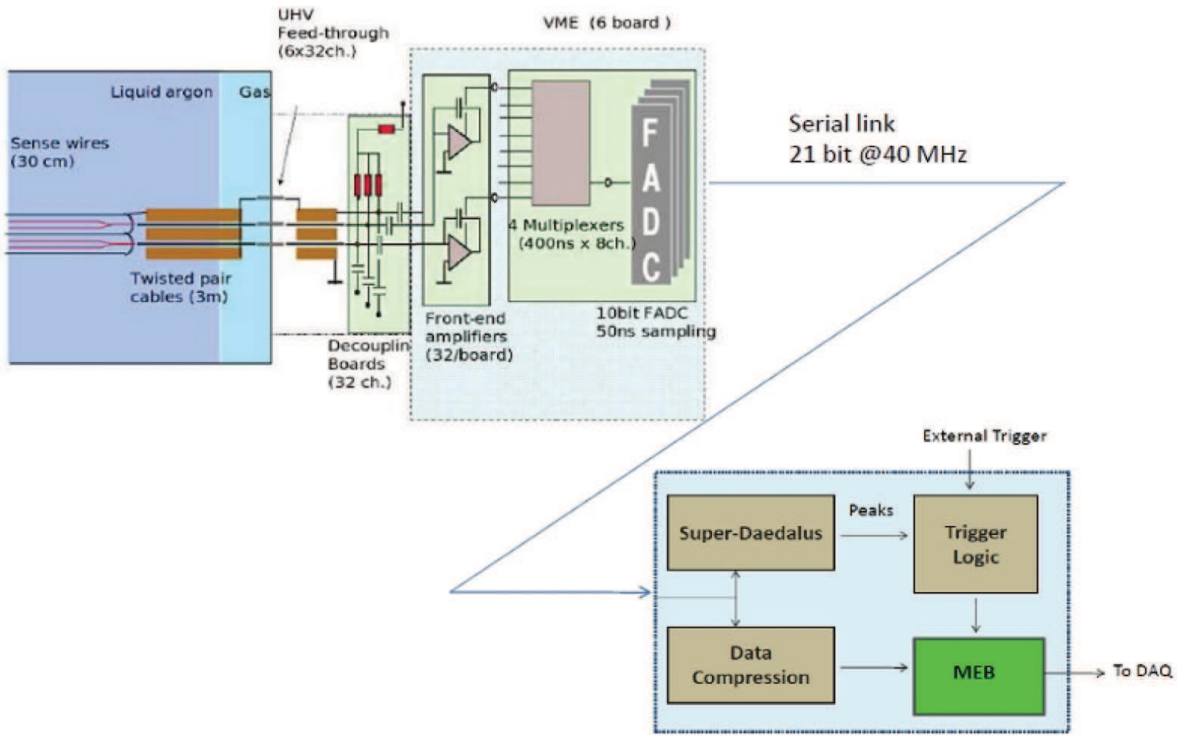}
\caption{Block diagram of the LAR-TPC electronic read-out.}
\label{block_readout}
\end{center} 
\end{figure}

 The read-out electronics is a subset of the one implemented in the ICARUS-T600 detector \cite{ICARUS-T600}.
 It's designed to provide continuous digitization and waveform recording of the signals from each TPC wire; 
 the chain is composed of three basic units, each one serving 32 channels: the Decoupling Board, the Analog Board, and the Digital Board.\\
 The Decoupling Board receives 32 analogue signals from the chamber and passes them to the analogue board via a decoupling 
 capacitors; it also provides wire biasing voltage and the distribution of the test signals.\\ 
 
 The Analogue Board hosts the front-end amplifiers and performs 16:1 channel multiplexing and 10-bit digitization at 40 MHz 
 rate (i.e. 400 ns sampling period per channel, corresponding to $\sim 0.6$  mm drift distance). The overall gain is around 5-6 ADC 
 counts/fC, i.e. about 1000 electrons per count, thus setting a minimum ionizing particle signal ($\sim 1$ fC/3mm) at about 10 ADC counts 
 with a dynamic range of about 100 mip's. To match the different signal's shapes two versions of this board have been used, differing
 for the integration time constant. The Collection version  has a time constant of 3 $\mu$s, much shorter than the signal duration, 
 and is used for  the unipolar signals coming from Collection wires; the area of the shaped signal is proportional 
 to the charge. The Induction version has a time constant of 100 $\mu$s, much longer than the signal duration, and is used for the bipolar 
 current  signals coming from the Induction wires.
 In the ICARUS  detector the Collection version is also used to instrument the Induction plane facing the drift volume, not instrumented in
 the present test-facility.

\begin{figure}[htbp]
\begin{center}
\includegraphics[width=10cm,angle=-90]{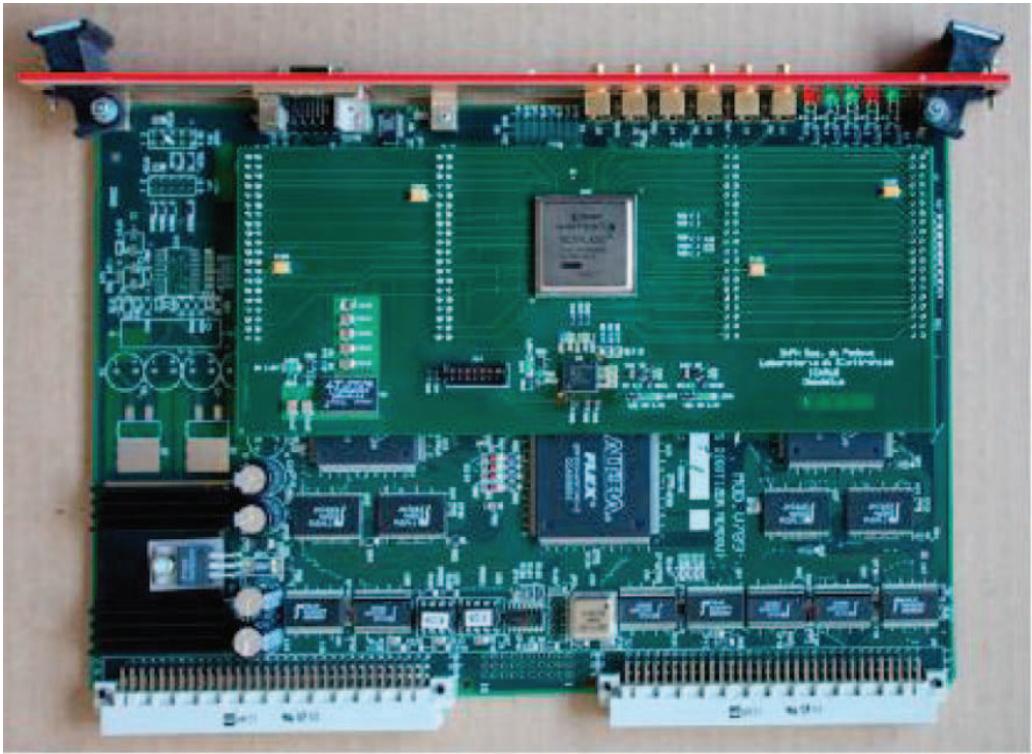}
\caption{Picture of the digital board equipped with the new SuperDaedalus chip, implementing the new hit-finding algorithm.}
\label{new_board}
\end{center}
\end{figure}

 The Digital Board (Arianna) acts as a 32 channels, 10 bit wide, waveform recorder. It continuously reads the data out of the V791 Board,
 stores them in multi event circular buffers (MEBs) and analyzes them according to a complex, programmable logic. MEBs length can be 
 chosen among 7 different values, ranging from 64 to 4096 t-samples (1 t-sample = 400 ns), corresponding to a drift distance from 38 
 mm to 2.5 m at nominal electric field; this feature permits a segmentation of data along the drift direction, defining  
 Regions of Interest (ROIs). As soon as a trigger signal is received, the active buffer is frozen, writing operations are moved to 
 the next free buffer, and the stored data are read out by the DAQ.

 Once the read out is completed, the frozen buffer is released and it's available to be overwritten by new incoming data. 
 This configuration guarantees no dead time, at least until the maximum DAQ throughput (up to 1 Hz, i.e. 1 full-drift event per wire per second in the
 ICARUS-T600 configuration) is reached.

 The buffer data are formatted in four different modes:

\begin{itemize}
 
\item RAW-DATA: no compression is performed, and data (Daedalus output + wire signal waveform) are written to the MEB as they are 
(Fig. \ref{Raw_data});
       
      \begin{figure}[htbp]
      \begin{center} 
      \includegraphics[width=2cm,angle=-90]{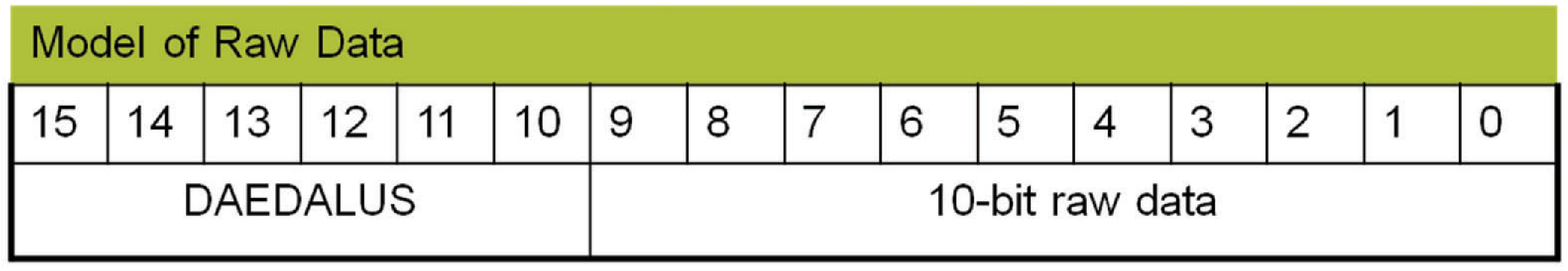}
      \caption{Raw data model of a 16 bit word.}
      \label{Raw_data}
      \end{center}
      \end{figure}

\item COMPRESSION 4: for each set of 4 wires, the signal amplitude value for a t-sample is compared to the previous one. If the difference 
is less than or equal to $\pm$ 7 ADC counts the four 4 bit differences are stored in a 16 bit word (Fig. \ref{Com4_Overfl}).
 If one or more differences are above that value (OVERFLOW) all the four differences are stored each one in a 16 bit word, flagged setting
  to ``1000'' the 4 more significant bits (``$1000 = 8$'', out of the compression 4 difference range); 
      
      \begin{figure}[htbp]
      \begin{center}
      \includegraphics[width=2.5cm,angle=-90]{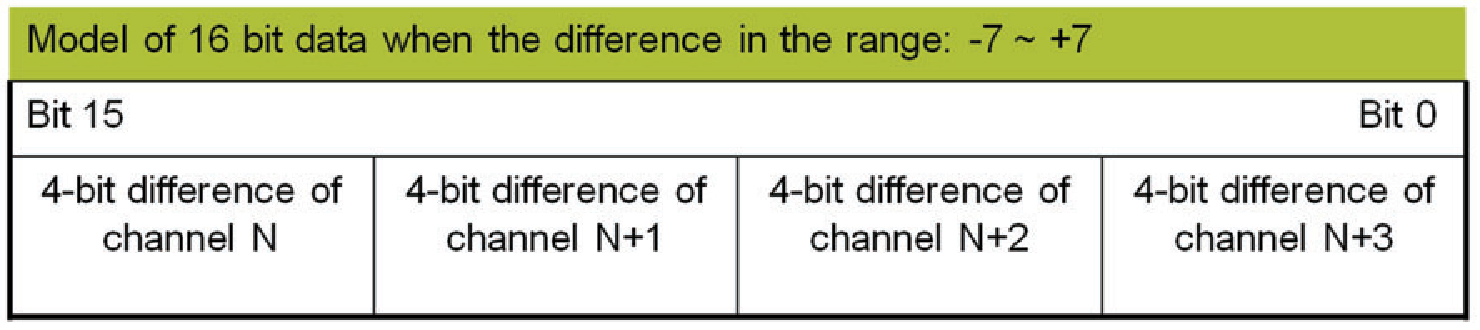}
      \includegraphics[width=2cm,angle=-90]{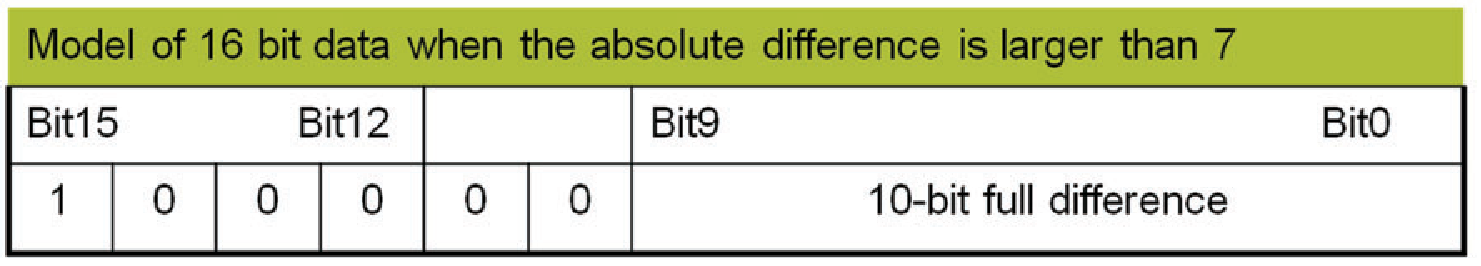}
      \caption{Compression  4 (top) and overflow (bottom) data model of a 16 bit word.}
      \label{Com4_Overfl}
      \end{center}
      \end{figure}

\item FULL-DIFFERENCE: for each wire the difference between two consecutive t-samples is written using 10 bit. Since the difference is 
preceded by the string "100000" this mode is equivalent to compression 4 during an overflow (Fig. \ref{Com4_Overfl}, bottom);

\item COMPRESSION 2: for each wire the difference is always written using 8 bit. For ``physical'' signals it never overflows (Fig. \ref{Comp_2}).
  
      \begin{figure}[htbp]
      \begin{center}
      \includegraphics[width=2.0cm,angle=-90]{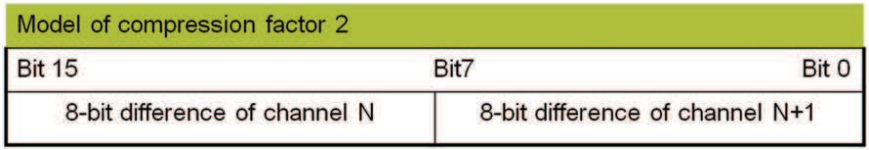}
      \caption{Compression 2 data model of a 16 bit word.}
      \label{Comp_2}
      \end{center}
      \end{figure}

\end{itemize}
 
 In addition in the ICARINO test-facility set-up both Induction and Collection boards are equipped with a new chip (SuperDaedalus)
 which implements a new hit-finding algorithm, here described in the following (Fig. \ref{new_board}).

 The test-run described here addressed the  study of the performance of the new digital boards with trough-going cosmic ray  muons both in
 terms of data compression, and ROI definition capability.
 The system was run with a global stop signal provided by a trigger system based on external scintillation counters or on the 
 internal TPC wires signal processed by the new SuperDaedalus  chip that produces a Global Trigger Output (GTO) from each digital board. 

\begin{figure}[htbp]
 \begin{center}
  \includegraphics[width=15cm]{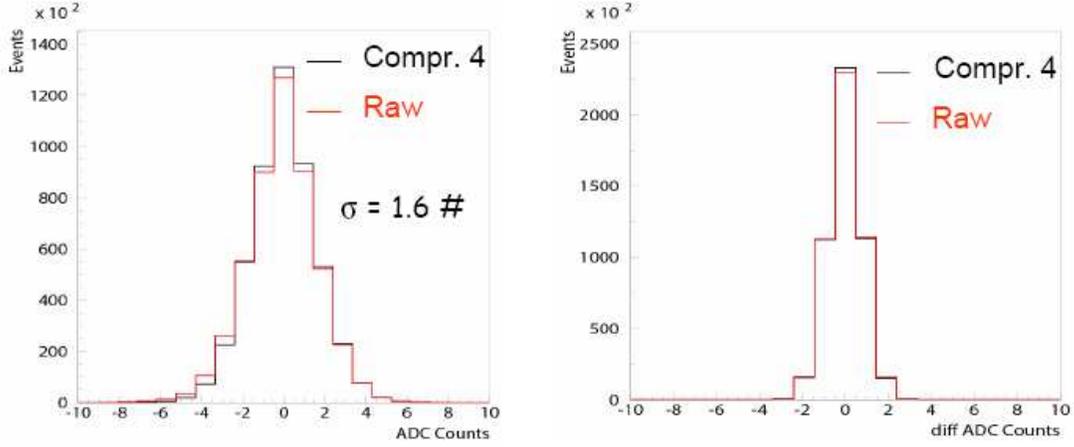}
 \end{center}
 \caption{Distribution of the wire signal pulse-height (left) and of the
          corresponding differences between two consecutive t-samples (right) for both
	  compression 4 (black) and raw (red) data modes (empty events).} 
 \label{fig:adc_fondo}
 \end{figure}

 All the data compression modes have been extensively tested in several runs. Thousands of test pulse and single cosmic muon events, both in raw data
 and compression 4 modes, were collected and analyzed to test the functionality of the compression  algorithm.
 No difference was found between the two modes by comparing the pulse-height and the difference between contiguous t-samples
 (Fig.\ref{fig:adc_fondo}).

 \begin{figure}[htbp]
 \begin{center}
 \includegraphics[width=13cm]{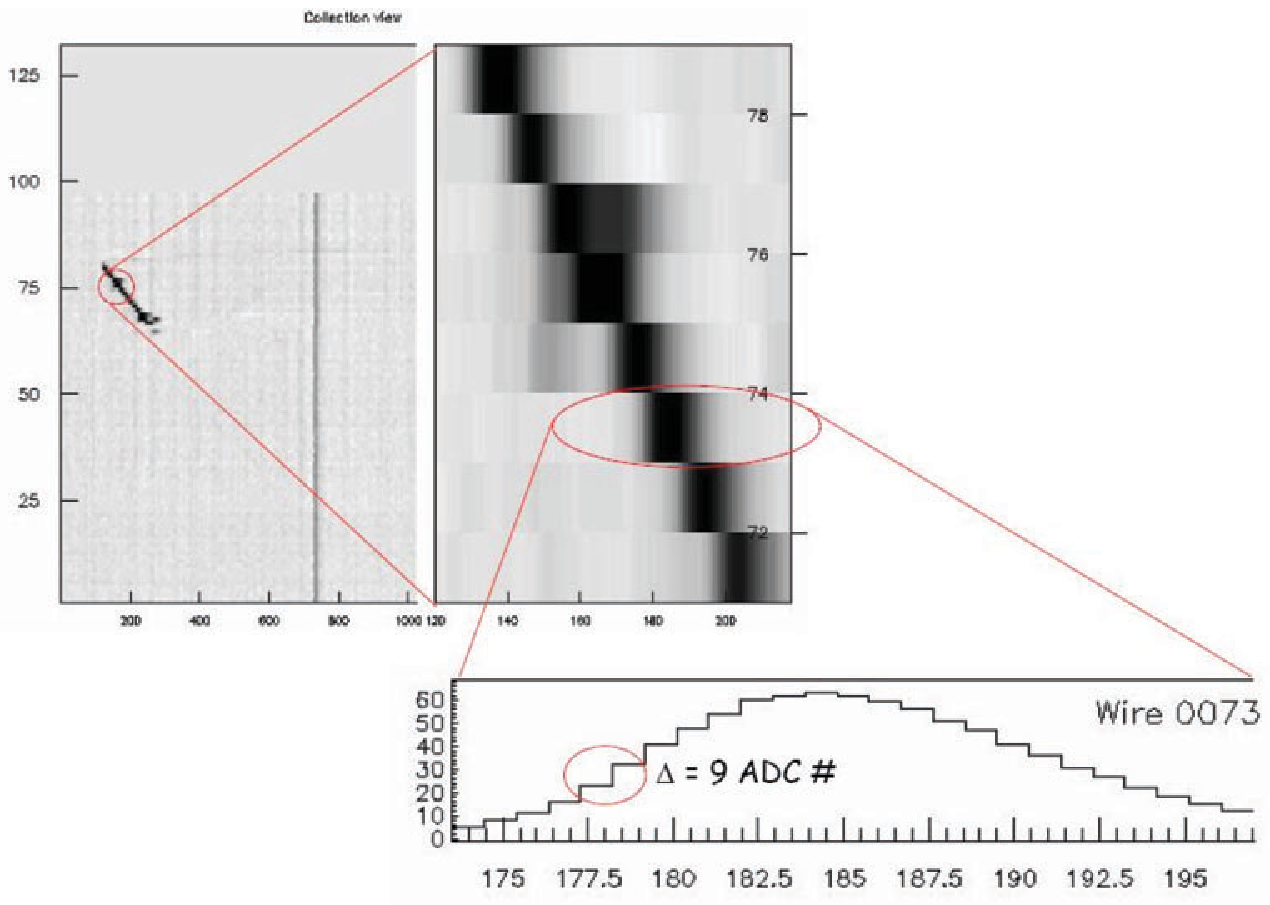}
 \includegraphics[width=13cm]{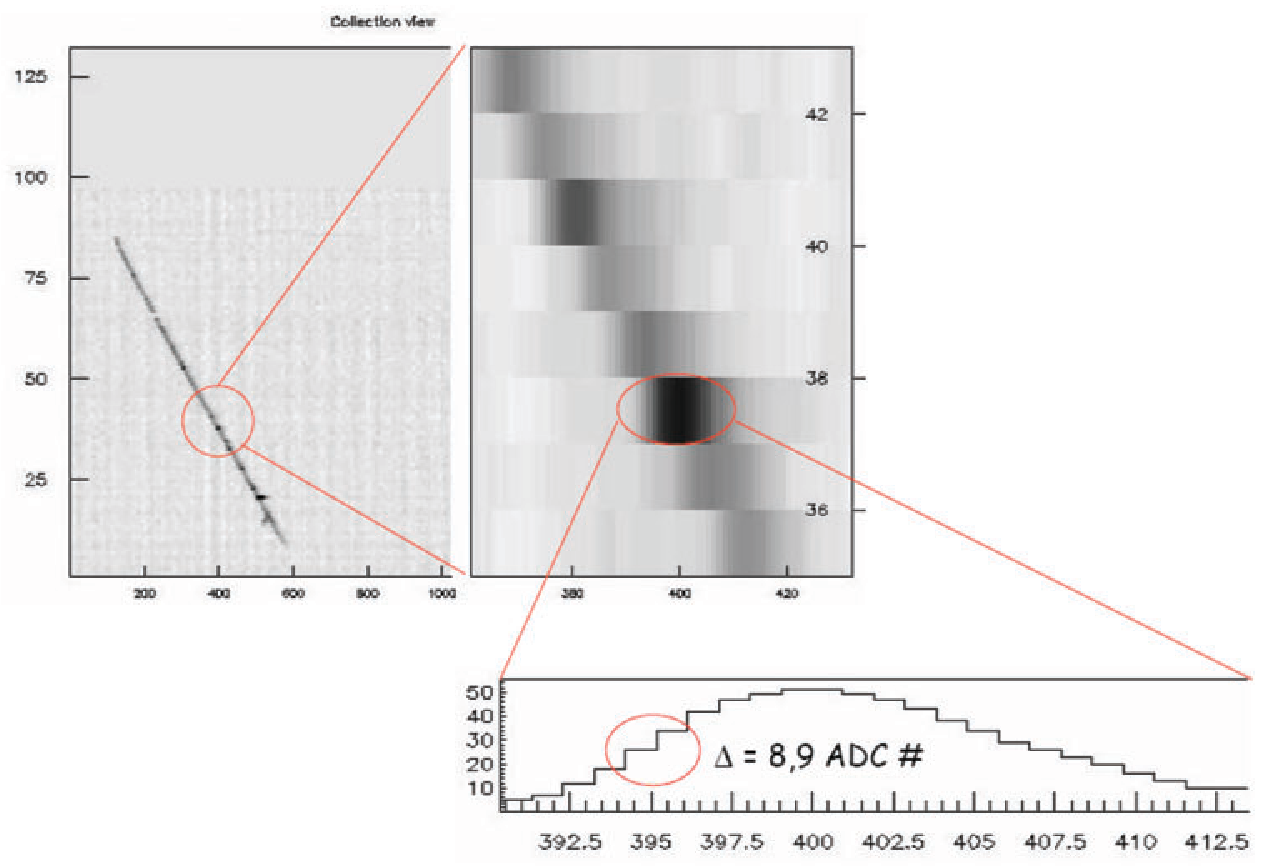}
 \end{center}
 \caption{Event with 1 (top) and 2(bottom) overflows correctly reconstructed
          within the Compression Mode 4.}
 \label{fig:overflows}
 \end{figure}

\begin{figure}[htbp]
\begin{center}
\includegraphics[width=8cm]{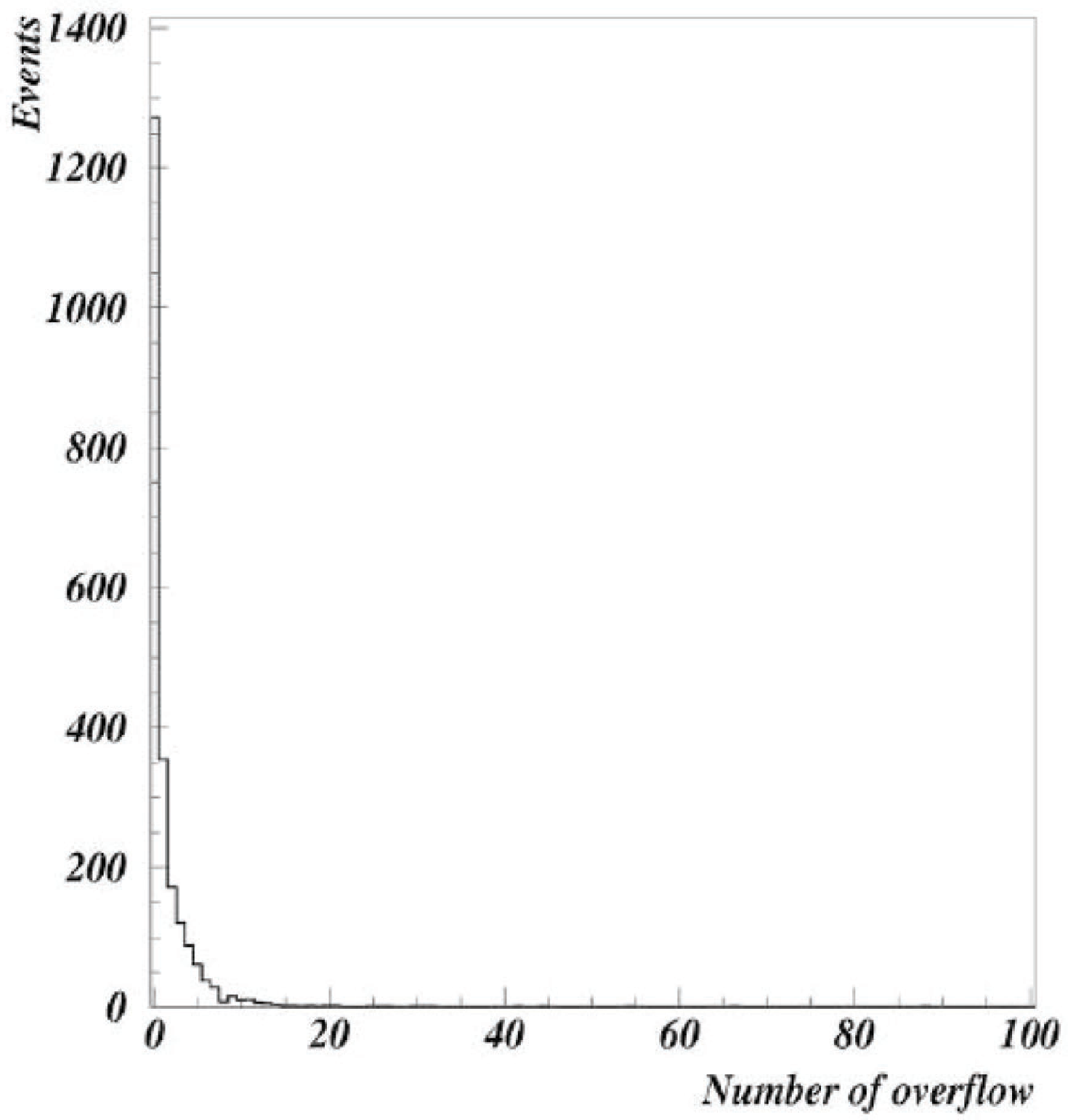}
\end{center}
\caption{The distribution of the number of overflows per event (Compression 
Mode 4).}
\label{fig:distribuzione_overflow}
\end{figure}

 About 30\% of the  events with single muon tracks resulted to be affected by  overflows  (Fig. \ref{fig:overflows}). The distribution of the number
 of overflows per event 
(Fig. \ref{fig:distribuzione_overflow}) shows on average 1 overflow per event. The detailed study of event reconstruction and signal shape
 in compression 4 mode next to the overflows shows no anomaly, no signal truncation or distortion.

\begin{figure}[htbp]
\begin{center}
\includegraphics[width=8cm]{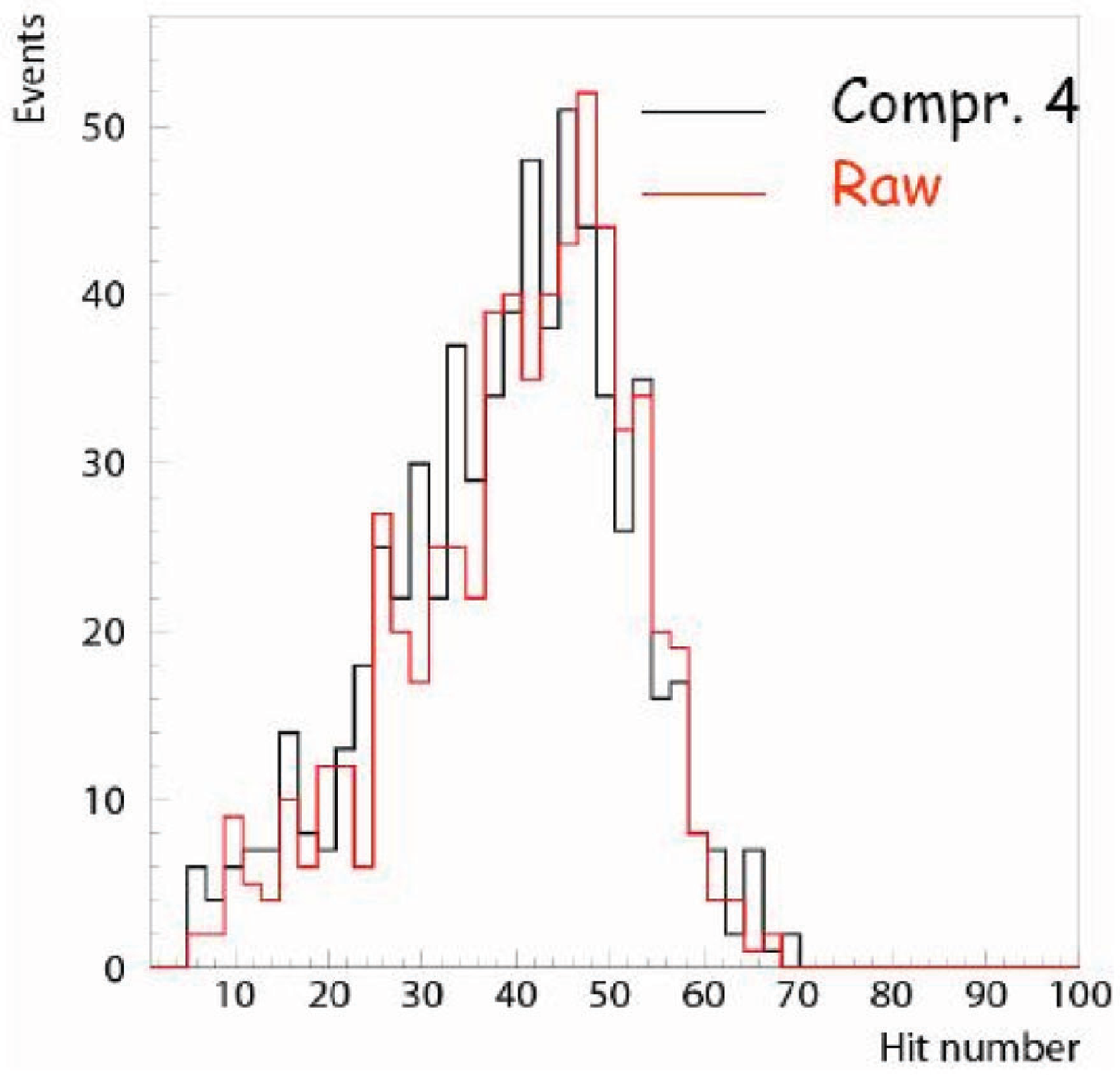}
\includegraphics[width=8cm]{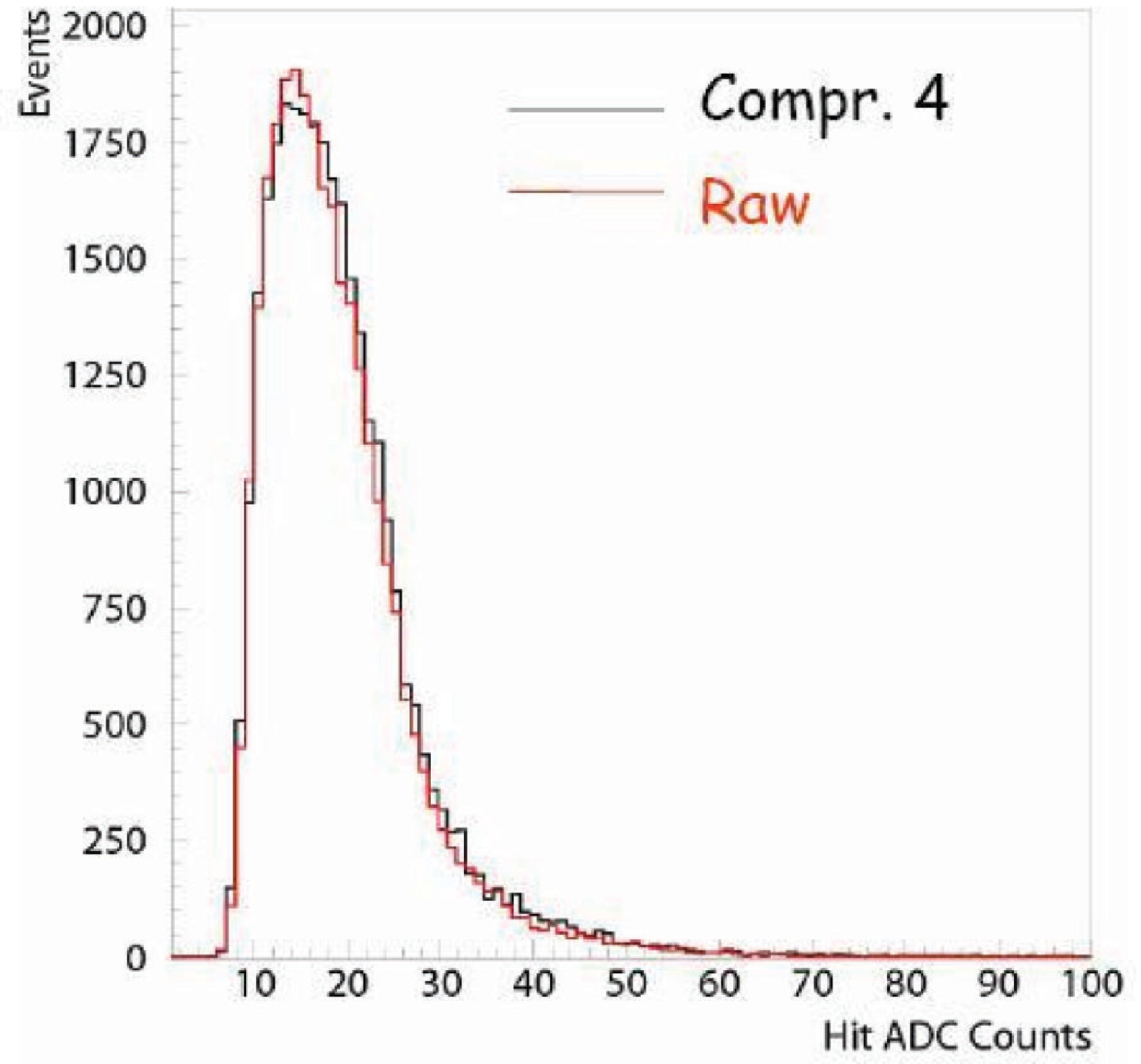}
\end{center}
\caption{Distribution of the hit number per track (top) and of their corresponding amplitude in ADC counts (bottom) in Compression 
Mode 4 and Raw data (muon events).}
\label{fig:distribuzione_hit_adc_eventi}
\end{figure}

The distributions of hits number and amplitude per track confirm that no difference is present between compression 4 mode and raw data 
(Fig.\ref{fig:distribuzione_hit_adc_eventi}), allowing to validate the new firmware of the digital boards. Therefore an effective 
reduction factor 3.9, with a consequent reduction in the read out time of the same factor, was established.

\section{The new hit finding algorithm: a trigger from the wire signals}

As observed during the 2001 test-run with a half of ICARUS T600 detector the hit signals on the Collection wires produced by a through-going 
muon are characterized by a pulse-height in the range 10$\div$30 ADC counts depending on the track inclination and extending over 
$\sim$ 25 t-samples. Data exhibit also the presence of a $\sim$ 4 ADC counts peak to peak high frequency noise with period less than 
10 t-samples overlapped to a 10 ADC counts low frequency oscillation of the baseline with a $\sim$ 1500 t-samples period.\\

 \begin{figure}[htbp]
 \begin{center}
 \includegraphics[width=12cm]{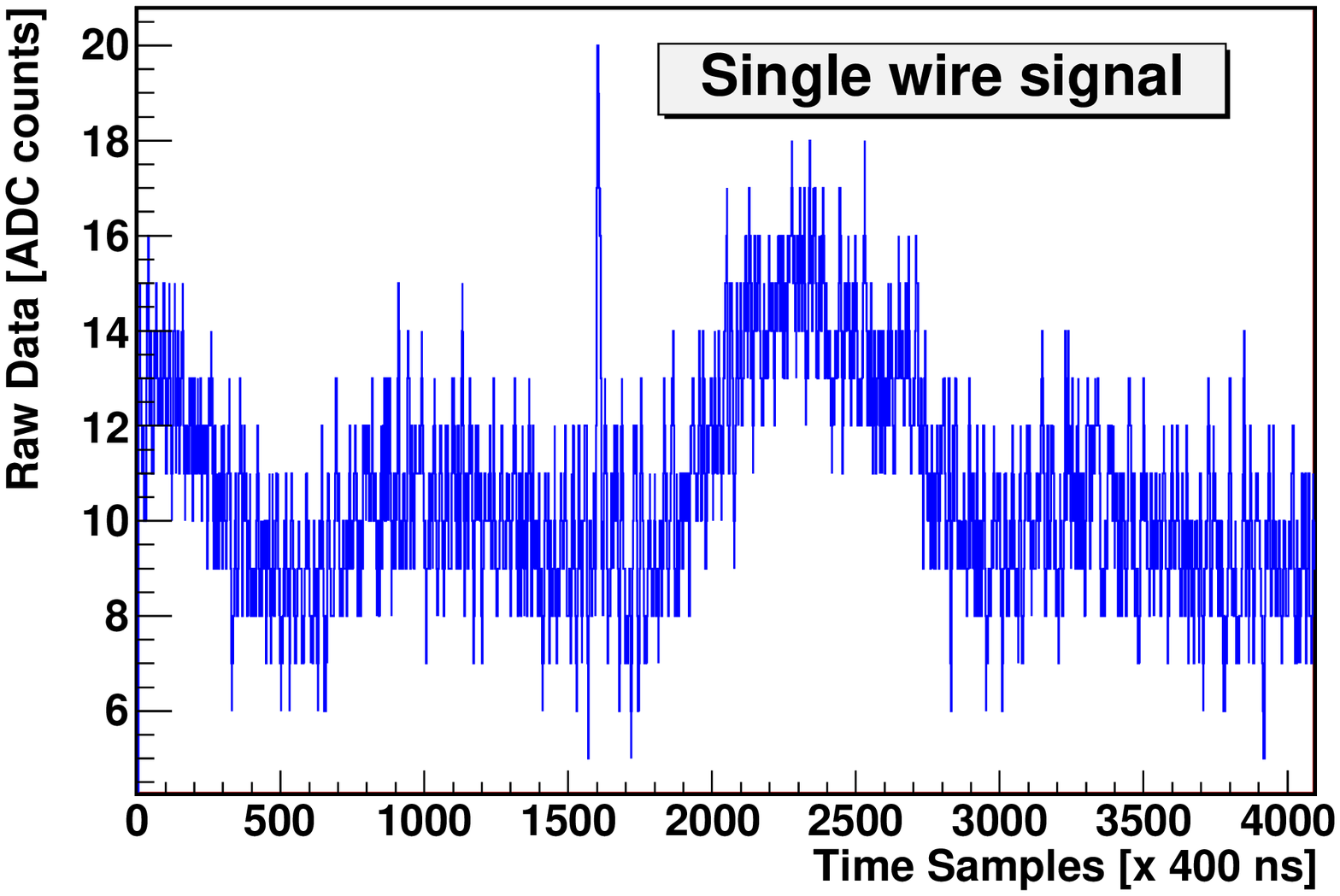}
 \caption{Typical signal shape of a Collection wire. Both the high and low frequency components
          of the electronic noise can be clearly seen.}
 \label{Pavia_wire}
 \end{center}
 \end{figure}
   
In order to extract the hit signal from the wire chambers, a dedicated algorithm had been developed with the aim of filtering out on-line 
at the same time both the low and high frequency components of the noise by means of a double rebinning technique applied to each wire 
signal. Since quick response time would be essential for trigger purposes, the algorithm was conceived as simple as possible to be easily 
implemented in a hardware frame. It consists in calculating the average of the signal amplitude Q(t) over a short $\sim 10$ t-samples and 
a long $\sim 250$ t-samples time interval, to treat the high and the low frequency components respectively:

\begin{equation}
\label{sr}
    Q_{short}(t_j;t_{j+10})=\sum_{i=j}^{j+10}Q(t_{i})/10  \quad \quad \quad
    Q_{long}(t_j;t_{j+250})=\sum_{i=j}^{j+250}Q(t_{i})/250.
\end{equation}

\noindent where the binning interval sizes have been initially chosen according to hit signal features
compared to the noise; the presence of the physical hit is then recognized requiring 
the difference $S(t)$ to exceed a fixed threshold $Q_{thr}$:

\begin{equation}
\label{dr}
    S(t) = Q_{short}(t) - Q_{long}(t) \ge \ Q_{thr}.
\end{equation}

\noindent
 The performance of this algorithm as well as the parameter choice was studied through 
 a software analysis of the cosmic muon data collected during the 2001 test run.
 As a initial result, the algorithm turned out full efficient ($\sim$ 100\%) in the physical hit recognition 
 with a negligible fake rate ($\le$ 3\%) for threshold values $Q_{thr} = 4 $ ADC
 counts over a wide range of the long rebinning interval ($100 \div 300$ t-sample).\\

 \begin{figure}[htbp]
 \begin{center}
 \includegraphics[width=12cm]{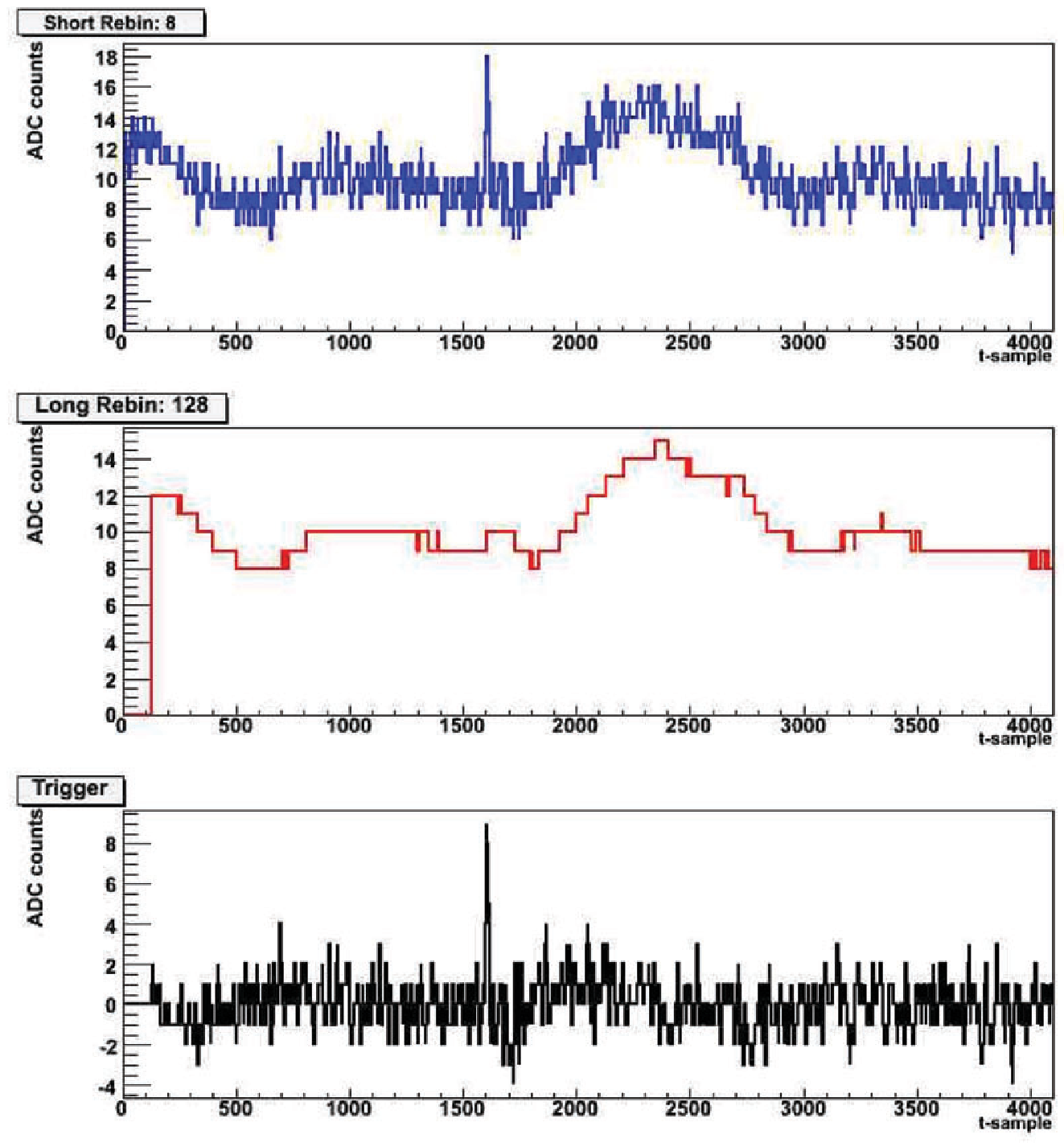}
 \caption{Example of the DR-slw algorithm applied to the wire signal in previous figure: 
          $Q_{short}(t)$ (top), $Q_{long}(t)$ (middle) and $S(t)$ (bottom).}
\label{doppio_rebinning_slw}
\end{center}
\end{figure}

  A slight modification with respect to the software algorithm was introduced in the hardware implementation
 of this simple algorithm. The interval into which the signal averages (long and short) are calculated is 
  a ``sliding window'' over the full data recording (DR-slw). Moreover new values of the binning interval sizes, 
  more suitable for the hardware implementation, 
 were chosen: 8 and 128 t-samples for the short and long binning respectively. 
 At each clock cycle the average over 8 and 128 t-sample is recalculated adding the value of the sample t0 and subtracting 
 that of sample t0-8 and t0-128 respectively.  When the difference of short and long rebinned signals is above a threshold 
 for at least 3 t-sample, a PEAK signal is output (fig. \ref{doppio_rebinning_slw}).

 \begin{figure}[htbp]
 \begin{center}
 \includegraphics[width=12cm]{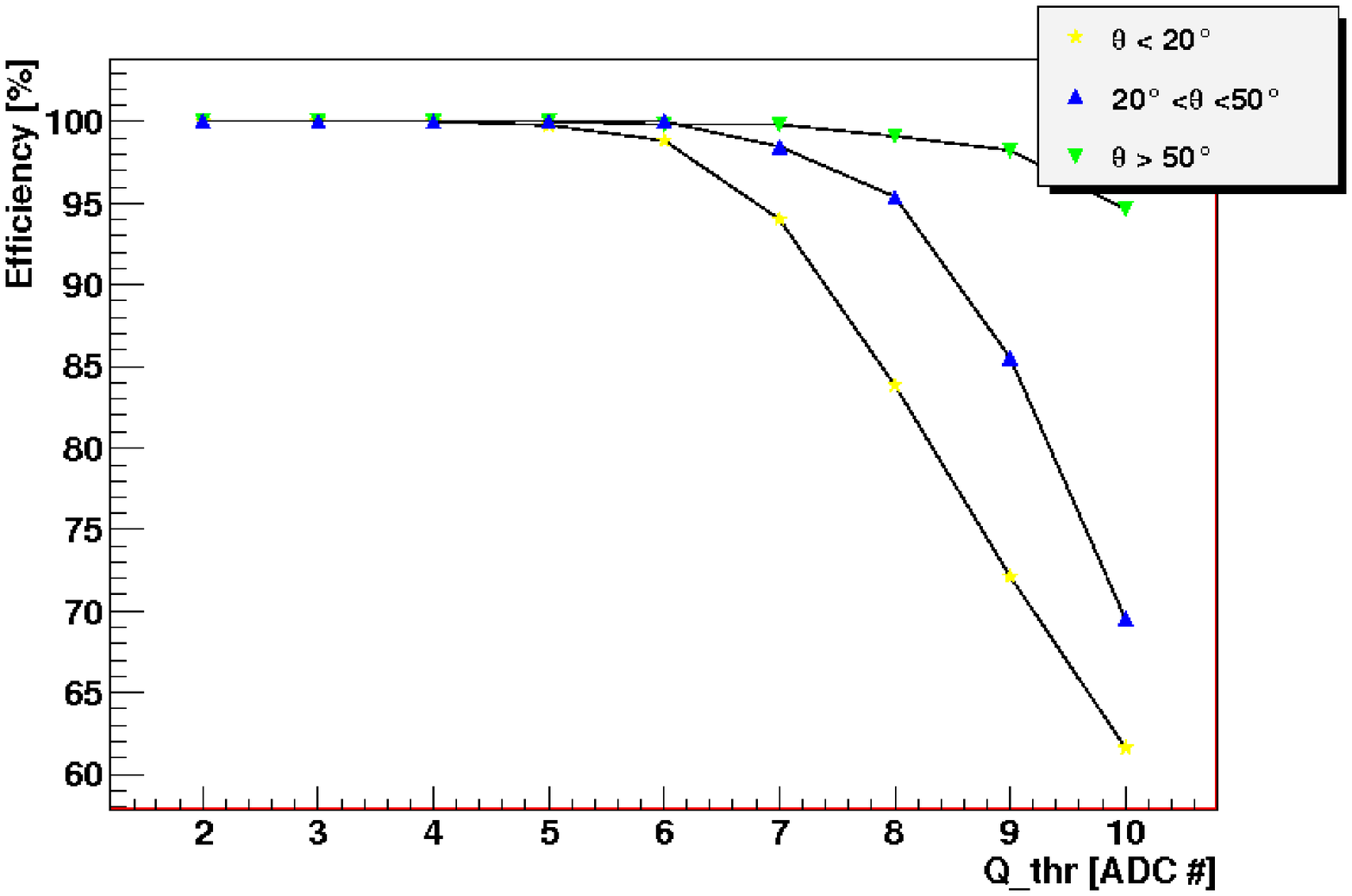}
  \includegraphics[width=12cm]{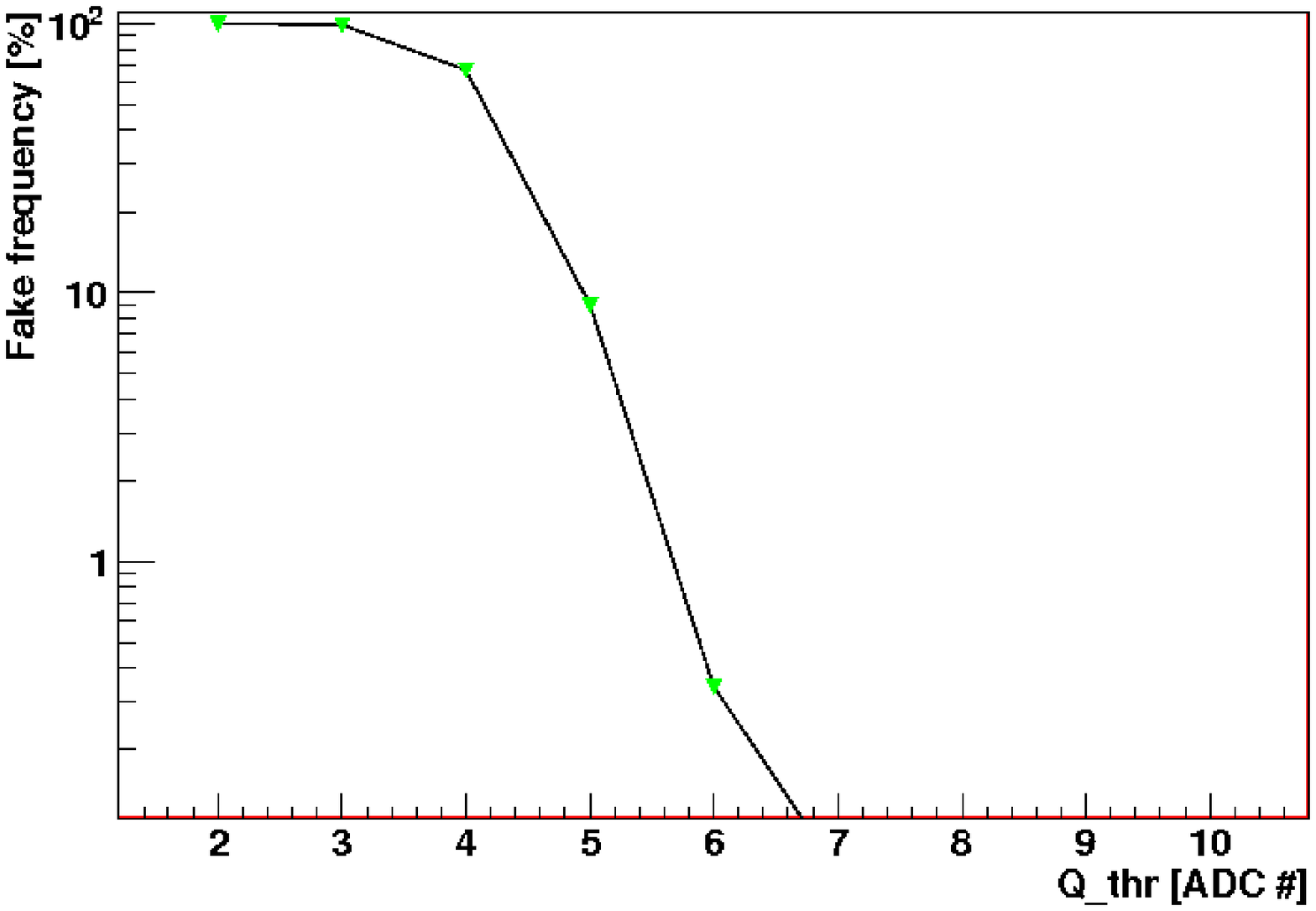}
 \caption{Hit identification efficiency for three different track inclination (top) and
  fake frequency (bottom) of the DR-slw algorithm as a function of the discrimination
  threshold in data collected in Pavia with the a semimodule of T600.}
\label{DRslw}
\end{center}
\end{figure}

\noindent
 Possible impact of these modifications on the algorithm performance has been 
investigated  on a sample of real muon events collected in Pavia. The result confirms the 
possibility to obtain full detection efficiency ($\sim$ 99\%) together with a negligible
fake signal rate ($\sim$ 1\%) applying a threshold value $Q_{thr}$ = 6 ADC counts, slightly higher than for 
the original algorithm (fig. \ref{DRslw}).

 \begin{figure}[htbp]
 \begin{center}
 \includegraphics[width=10cm,angle=-90]{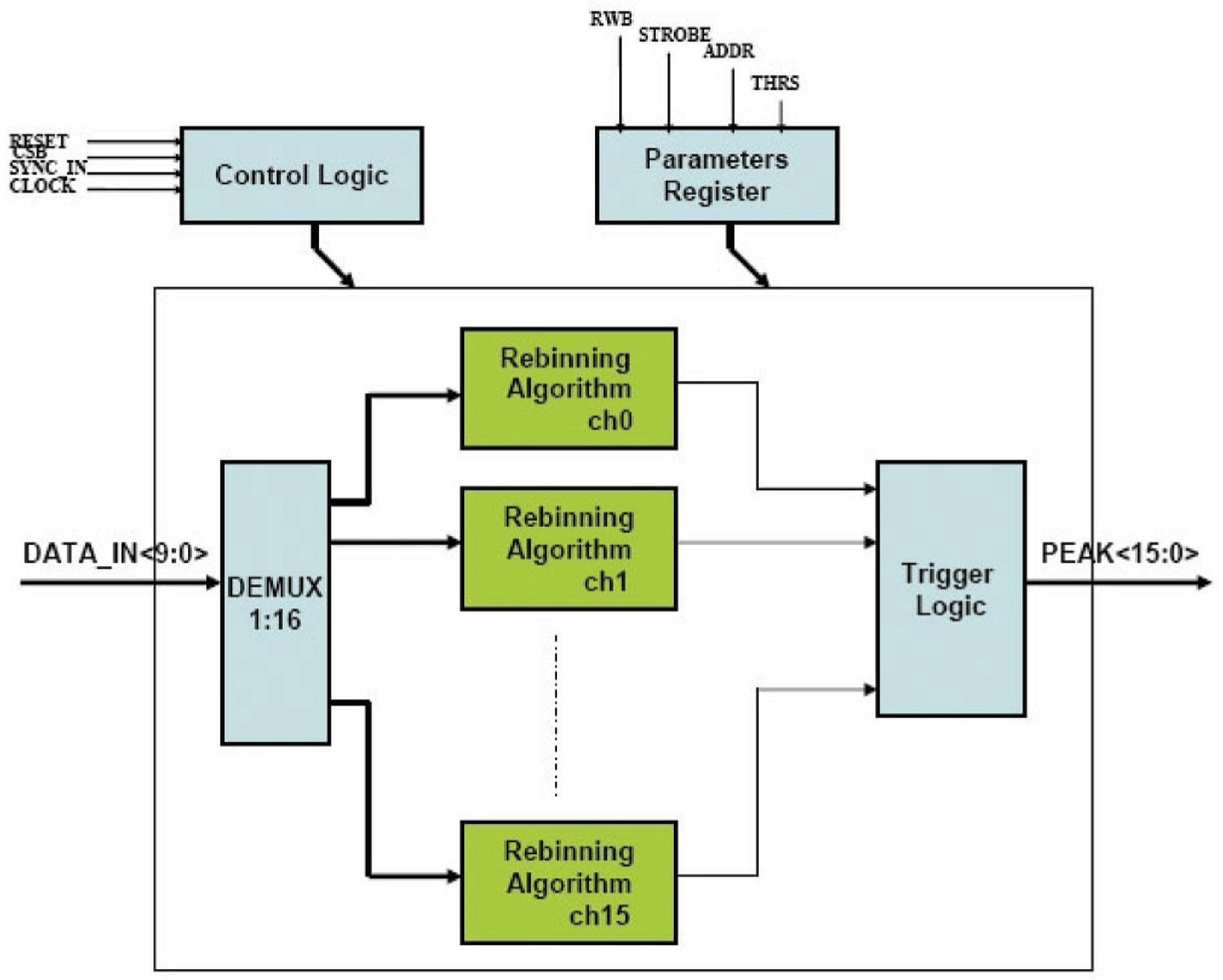}
 \caption{Block diagram of the DR-sw hit finding procedure and generation of a local trigger signal.}
 \label{block_trigger}
 \end{center}
 \end{figure}

 The hit identification procedure can also be used for the recognition of a ROI and the generation of an internal trigger.
 The block diagram of the on-line algorithm implementation is shown in Fig. \ref{block_trigger}. 
 Serial data (10 bit for each channel) are de-multiplexed to enter the 
 rebinning algorithm. Through the parameter "polarity" it's possible to set the trigger
 on the rising or on the falling edge. In particular the falling edge is used for the Induction wires, due to the presence
 of an undershoot on those signals.
 
 To get a trigger from the PEAK signals a majority stage has been included. Dividing the 32 wires in groups of 16, and taking
 the logical OR of the two majority, the rate of fake triggers can be reduced, while keeping the system sensitive to the 
 short tracks (16 wires correspond to 5 cm). A possible drawback of this solution could arise for inclined tracks: 
 as the PEAK signals may not overlap, the efficiency of the majority would decrease with the angle (Fig. \ref{DRslw}) . To avoid 
 this situation, the final stage of the SuperDaedalus performs a stretching of the PEAK signals among four values, 25, 25, 75, and 125 $\mu$s;
 analysis made on data of the 2001 test-run showed that  the efficiency never fall below $99\%$ for traces of all angles if
 a stretchng of 50 or 75 $\mu$ s is applied.

 In order to determine the optimal parameters for the new Collection boards,
 cosmic muons track events have been collected with the ICARINO test-facility with the external scintillation
 counter trigger. Then the new GTO internal signal from Collection boards has been used to 
 trigger cosmic muons and to study the performance in Induction view also.
 
 Finally a particular configuration of internal TPC trigger, aiming to the detection of isolated low 
 energy events (solar neutrino like) 
 has been set-up. This last measurement represent a crucial test in view of the future development 
 of huge LAr-TPC detectors, since the localization of low energy events represents a challenging 
 item for large volume detectors.

\begin{figure}[!htb]
\begin{center}
\includegraphics[width=\textwidth]{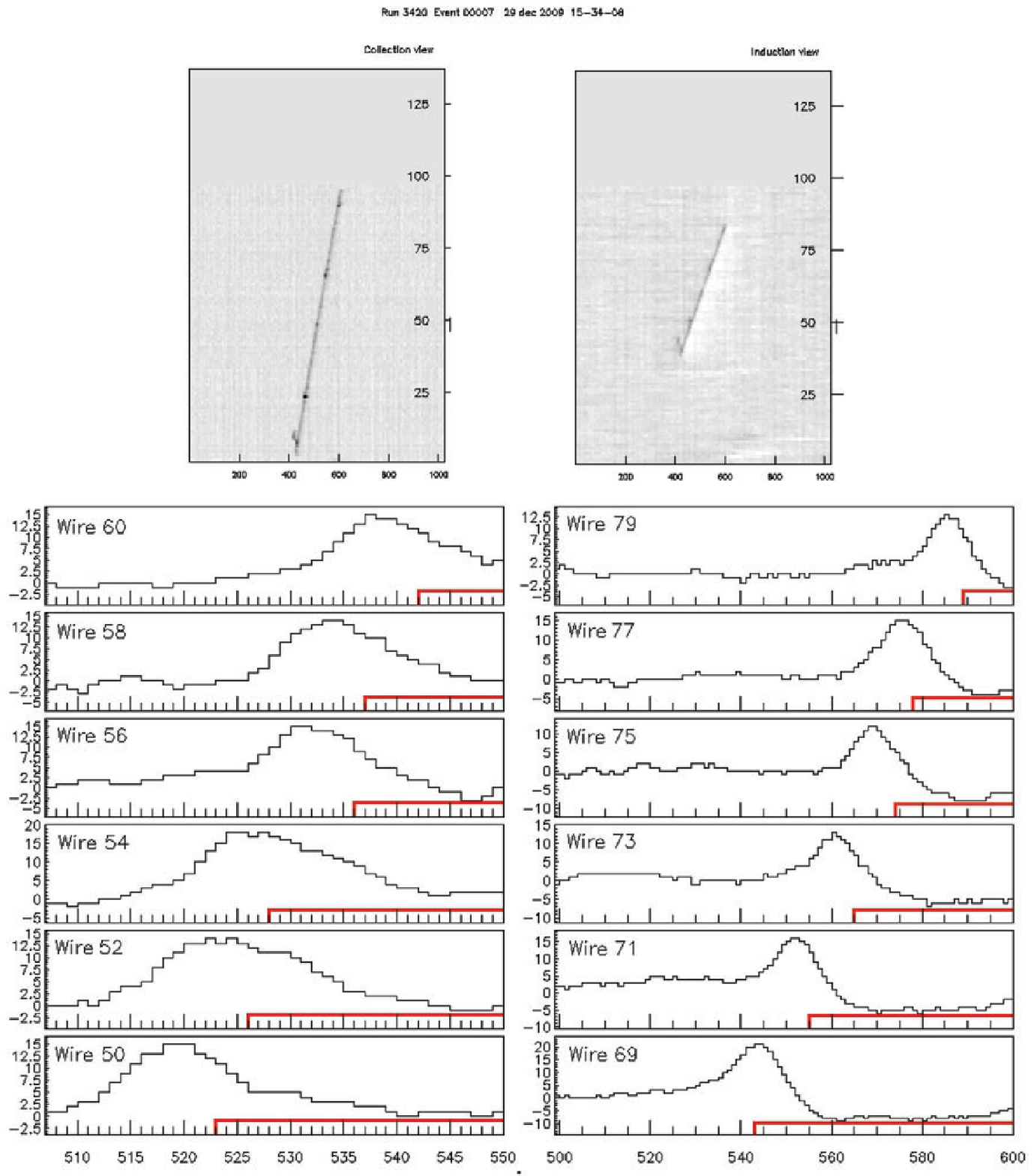}
\end{center}
\caption{Visualization of an inclined muon track in the wire/t-sample plane. The hit signals in
 the Collection (left) and Induction (right) view are drawn as a function of the drift coordinate;
  the red line represents the peak signal for the threshold value $Q_{thr}$= 6 $\#$ ADC.}
\label{fig:event_peak}
\end{figure}

\begin{figure}[!htb]
\begin{center}
\includegraphics[width=1.15\textwidth]{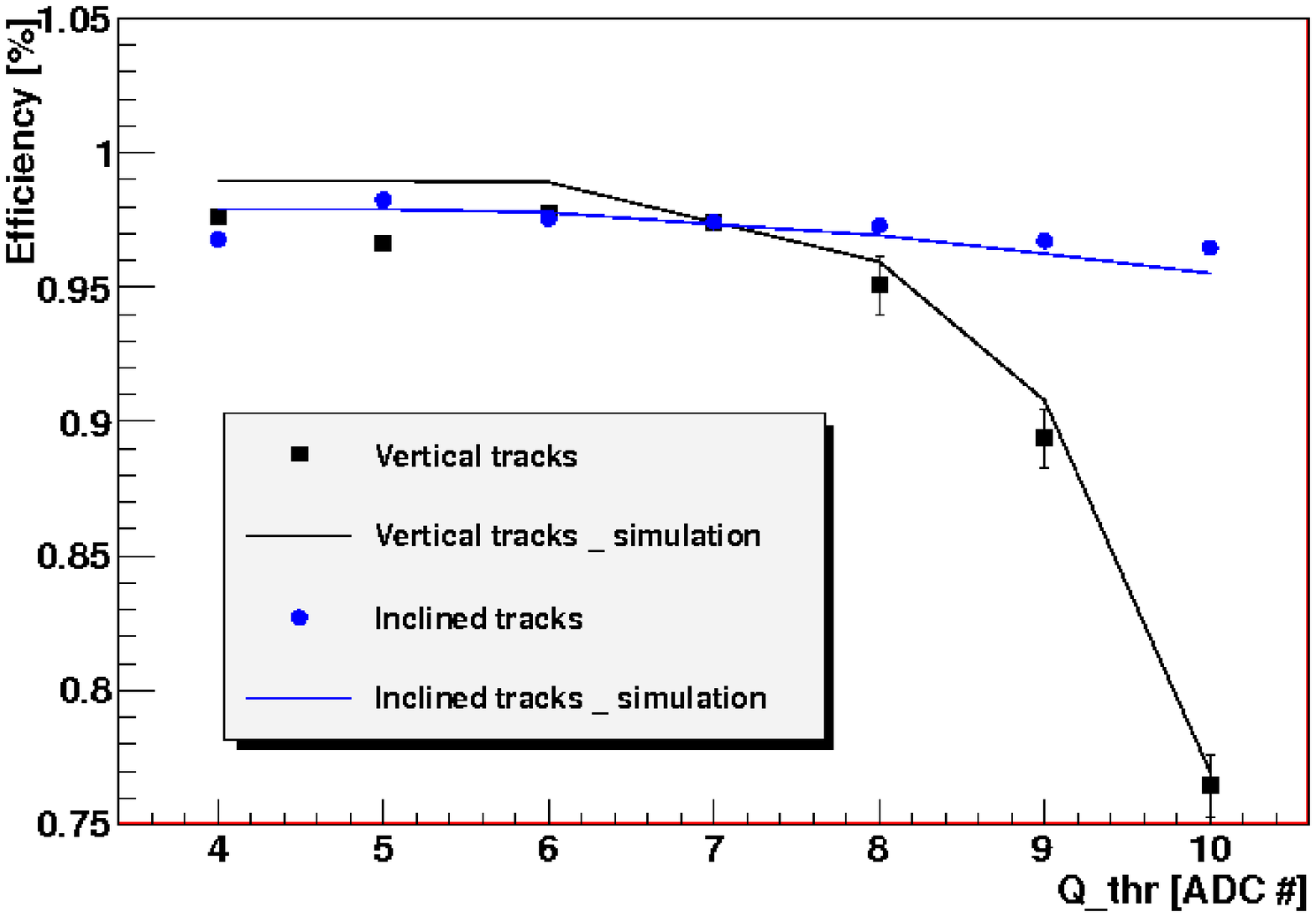}
\end{center}
\caption{Mean hit identification efficiency on the single wire in Collection view as a 
function of the threshold value $Q_{thr}$  measured for vertical 
and 45$^{\circ}$ inclined muon tracks. As a reference the values obtained with
  the software simulation of the DR-slw algorithm are also represented.}
\label{fig:eff_filo}
\end{figure}

\subsection{Single hit detection in Collection view}

\noindent
 The hit finding efficiency on the single wire in Collection view has been measured as a 
 function of the threshold value $Q_{thr}$ ranging in the interval 4$\div$ 10 \# ADC, with the 
 stretching parameter set at 50 and 75 $\mu$s and requiring the coincidence within 
 10 $\mu$s with the hit identified in the off-line analysis.
  Among all the vertical and 45$^{\circ}$ inclined muon tracks, collected by means of the external trigger, only the through
 going muons without energetic $\delta$ rays were selected (fig. \ref{fig:event_peak}).
 The statistic was  $\sim$ 1000 and $\sim$ 500 events for vertical and inclined tracks respectively.
  
 As a result an almost full single hit finding efficiency was measured for different track inclinations
 at low $Q_{thr}$ which degrades with  increasing the $Q_{thr}$ threshold. Moreover efficiency  values above 95 \% 
 wwere measured for $Q_{thr}\leq 7 $ ADC counts independently of the dip angle (fig. \ref{fig:eff_filo}).

\begin{figure}[!htb]
\begin{center}
\includegraphics[width=0.9\textwidth]{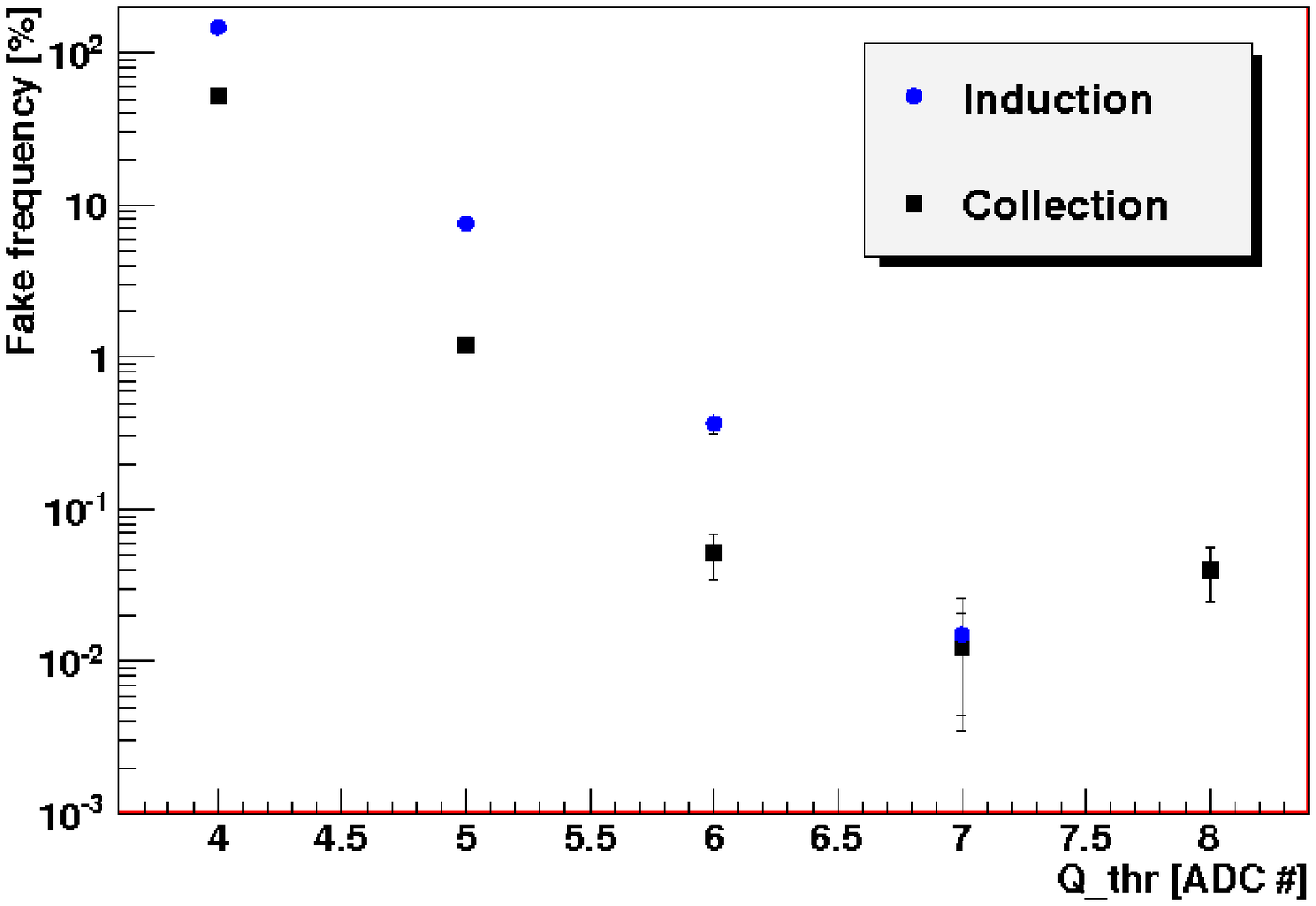}
\end{center}
\caption{Average fake trigger rate per event on the single wire in Collection  and Induction
 view, measured on the whole 1024 t-samples interval.}
\label{fig:fake_filo}
\end{figure}

\noindent
 A dedicated run with random external trigger was performed to measure the corresponding fake hit detection
 rate on the single wire recognizing the presence of the PEAK signals on the whole 1024 t-samples interval 
 in empty events (i.e. with no hits identified by the off-line analysis software). The rate of fakes per event
 was rapidely decreasing down to $10^{-3}$  with threshold  for both Collection and Induction views for
  $Q_{thr} \geq 6  \#$ ADC (fig.\ref{fig:fake_filo}) (1000 events overall collected statistics).
 The obtained results demostrated the single hit finding reliability, allowing to work satisfactory, with high efficiency
 and low associated fake hit signals.

\begin{figure}[tb]
\begin{center}
\includegraphics[width=0.8\textwidth]{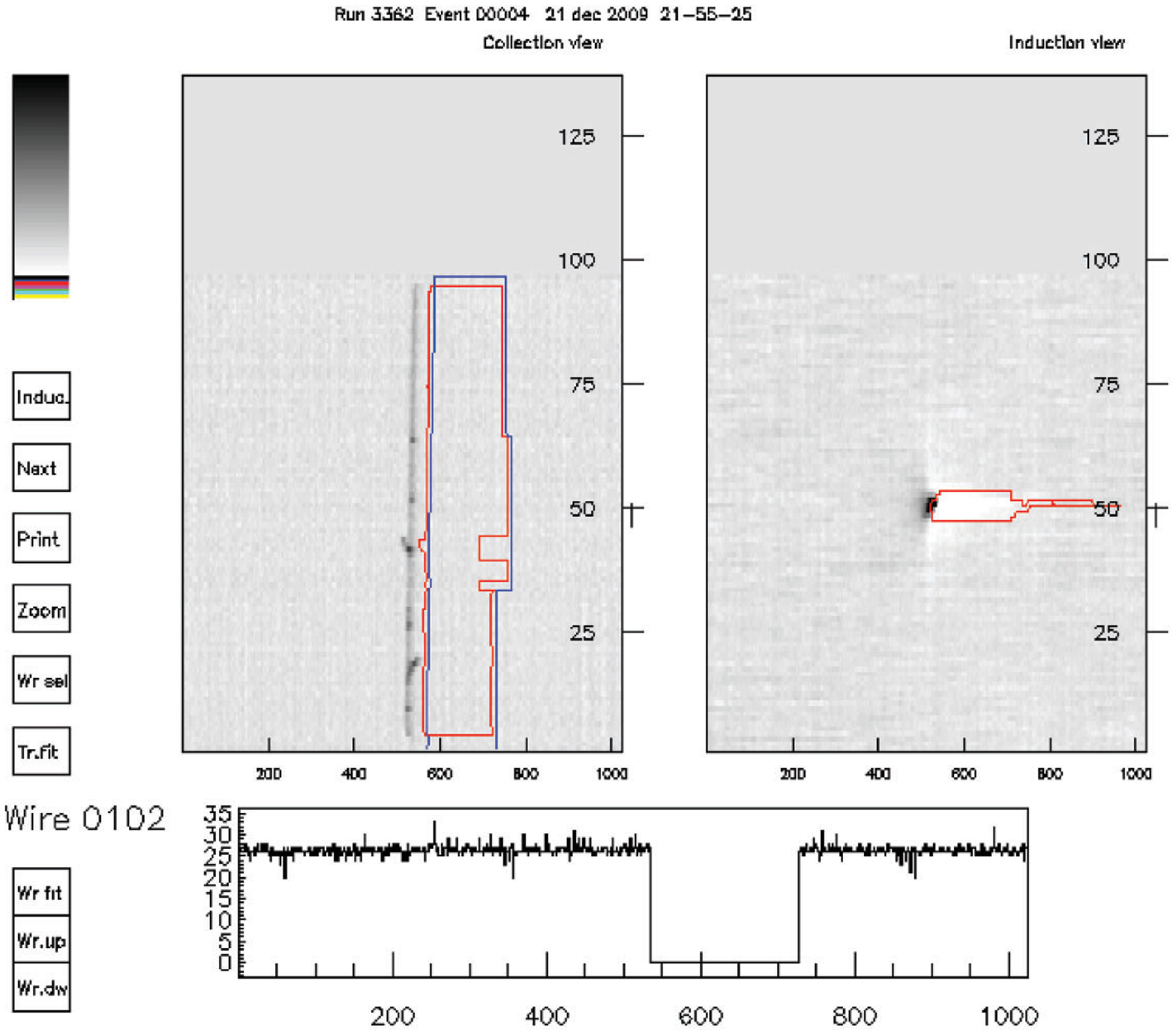}
\end{center}
\caption{Visualization of a vertical through-going muon track in the wire/t-sample plane, i.e. parallel
 to the TPC wire planes, both in Collection and Induction views. The red contour define the region 
 in which a peak signal is present for a threshold value $Q_{thr} = 6 \#$ ADC, while the blue 
 line indicates the GTO signal for a Majority value $M = 8$. The profile of a typical GTO signal 
 recorded in Collection view is also shown (bottom).}
\label{fig:event_display_peak_maj}
\end{figure}

\subsection{Study of the GTO signal in Collection view}
 
\noindent 
 Characteristics and performance of the GTO signal in Collection view were studied on
 a sample of clean through-going muons tracks, selected among vertical and 45$^{\circ}$  
 cosmic muon events collected with the external trigger system (fig. \ref{fig:event_display_peak_maj}). 
 GTO signals from two out of the three Collection boards were recorded for a fixed threshold 
 value $Q_{thr} = 6$ (chosen relying on the analysis described in the previous section), for
  different values of the majority ($M = 8, \, 12, 15$) and stretching parameters.

\begin{figure}[tb]
\begin{center}
\includegraphics[width=0.9\textwidth]{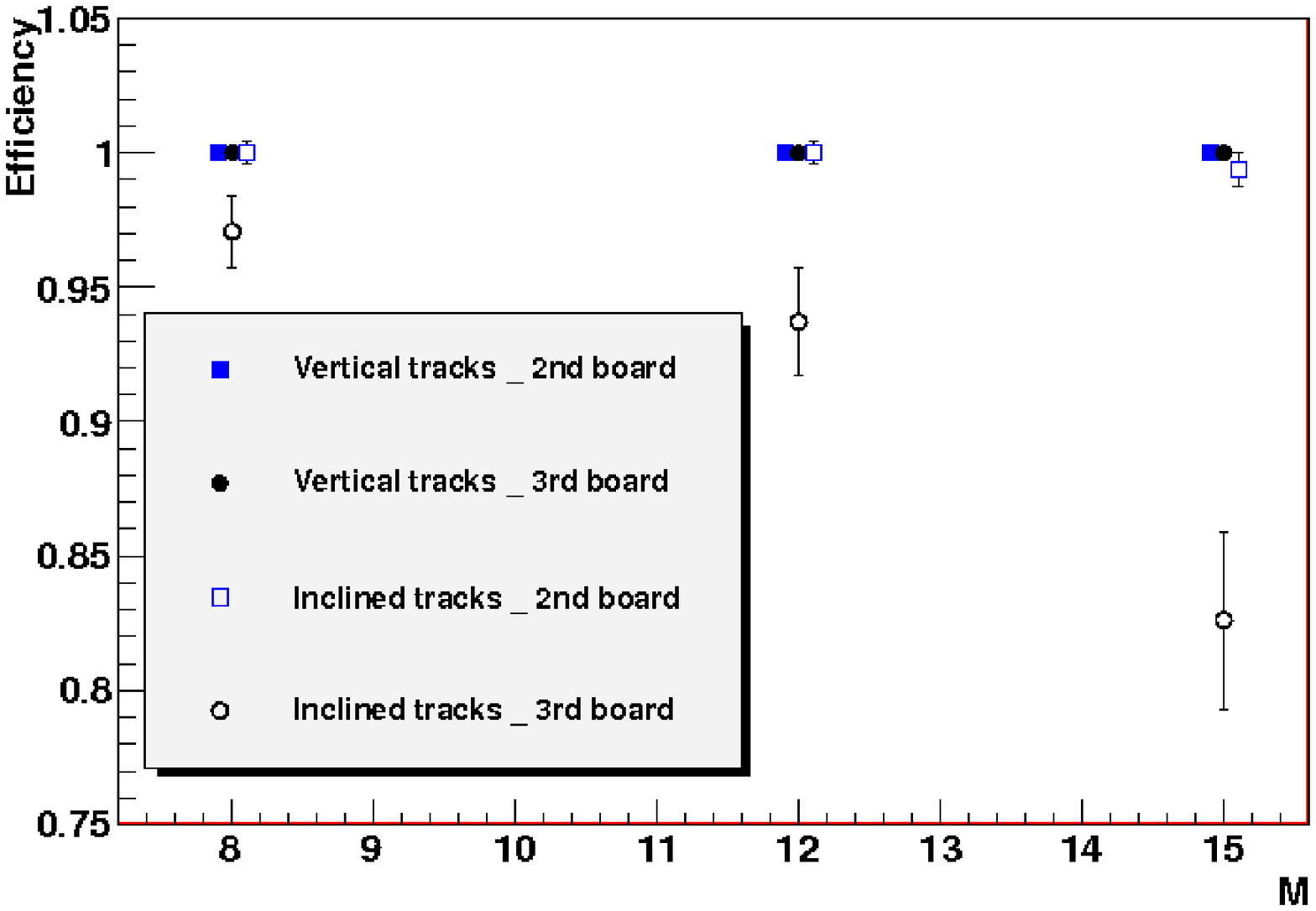}
\end{center}
\caption{Efficiency of the GTO signal per event in Collection view for the smallest value 
of the stretching parameter, $25 \div 50 \mu$s, and $M = 8,\, 12,\, 15$, for vertical 
and $45^0$  tracks as measured in the second  and third  board.}
\label{fig:Maj_Coll_eff_short}
\end{figure}

\noindent

 Time delay of the GTO signal with respect to the first track hit,
 depends mainly on the track dip angle due to time displacement of the PEAK signals on consecutive wires and on the required
 majority level. Therefore the GTO signal efficiency was determined into a suitable time window 
 of 50 and 300 t-samples depending of  the track dip. 
 An almost 100 $\%$ efficiency has been observed for majority values up to $M = 15$ for inclined and vertical tracks for both 
 boards (fig. \ref{fig:Maj_Coll_eff_short}) even with the shortest stretching parameter. 
 Actually small inefficiencies for in the first board for the inclined tracks sample are caused by 
 events not completely contained, entering the TPC 
 volume after the 24$^{th}$ wire and then with less than 8 hits on the first board.

\begin{figure}[!htb]
\begin{center}
\includegraphics[width=\textwidth]{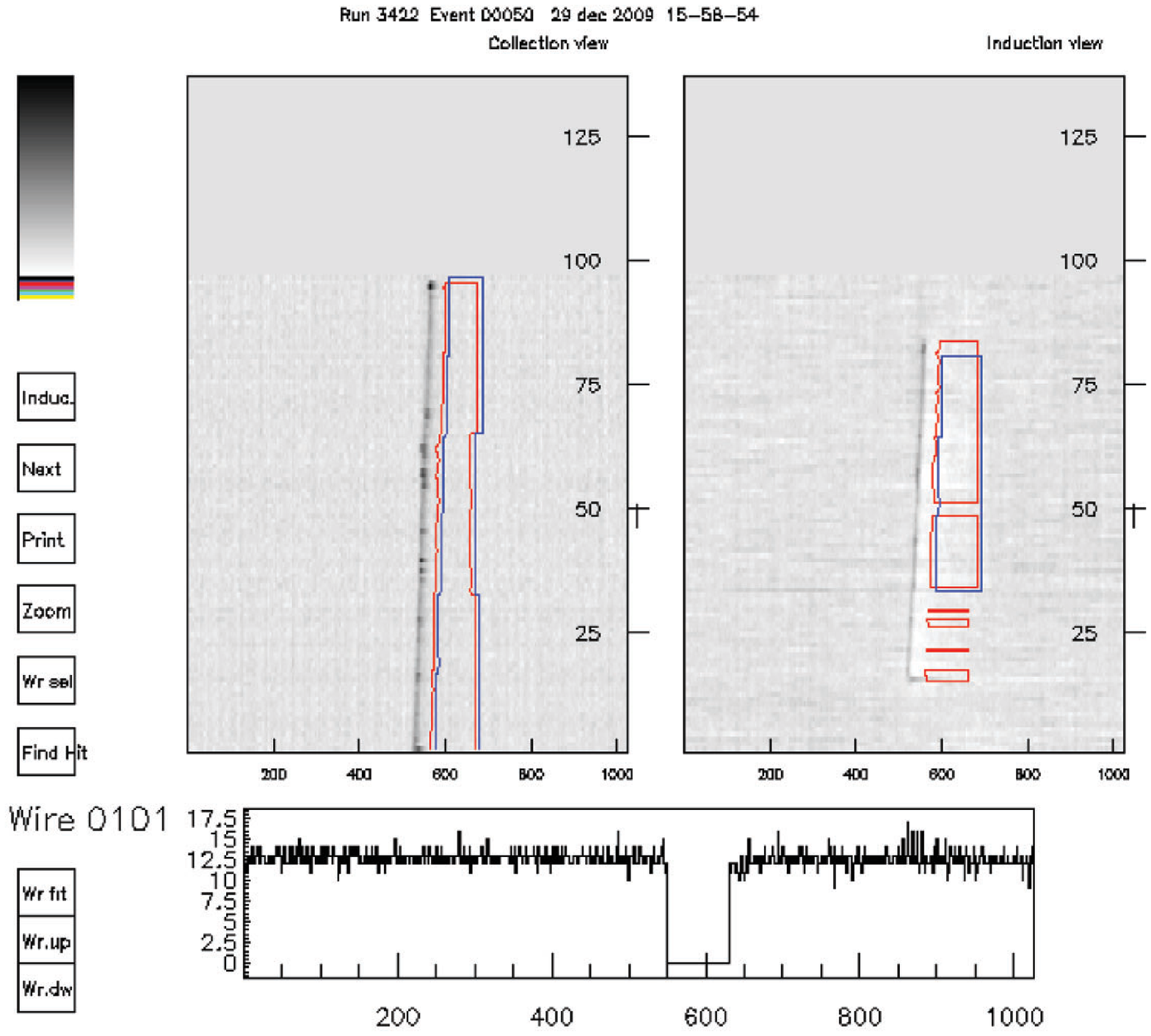}
\end{center}
\caption{Visualization of a through-going muon track parallel to the TPC wire planes and with
 $45^{\circ} \ \pm \ 5^{\circ}$ inclination with respect to the Induction and Collection wires
  orientation. The values of the threshold and majority parameters are 
  $Q_{thr} = 6 \#$ ADC and $M = 8$ respectively.}
\label{fig:event_display_peak_maj_Ind}
\end{figure}

\subsection{Internal trigger based on the Collection GTO signal to search for tracks in Induction view}

\noindent
 An internal trigger system based on the coincidence of the GTO signals from the
 first and the third boards in Collection view was set-up. A sample of events of through going 
 $\mu$ almost parallel to the TPC wire planes were collected and  used 
 to study the performance of the peak  signal in Induction view. In particular, only 
 tracks with $45^{\circ} \ \pm \ 5^{\circ}$ inclination with respect to the Induction and
  Collection wire orientation were selected (fig. \ref{fig:event_display_peak_maj_Ind}).

\begin{figure}[!htb]
\begin{center}
\includegraphics[width=\textwidth]{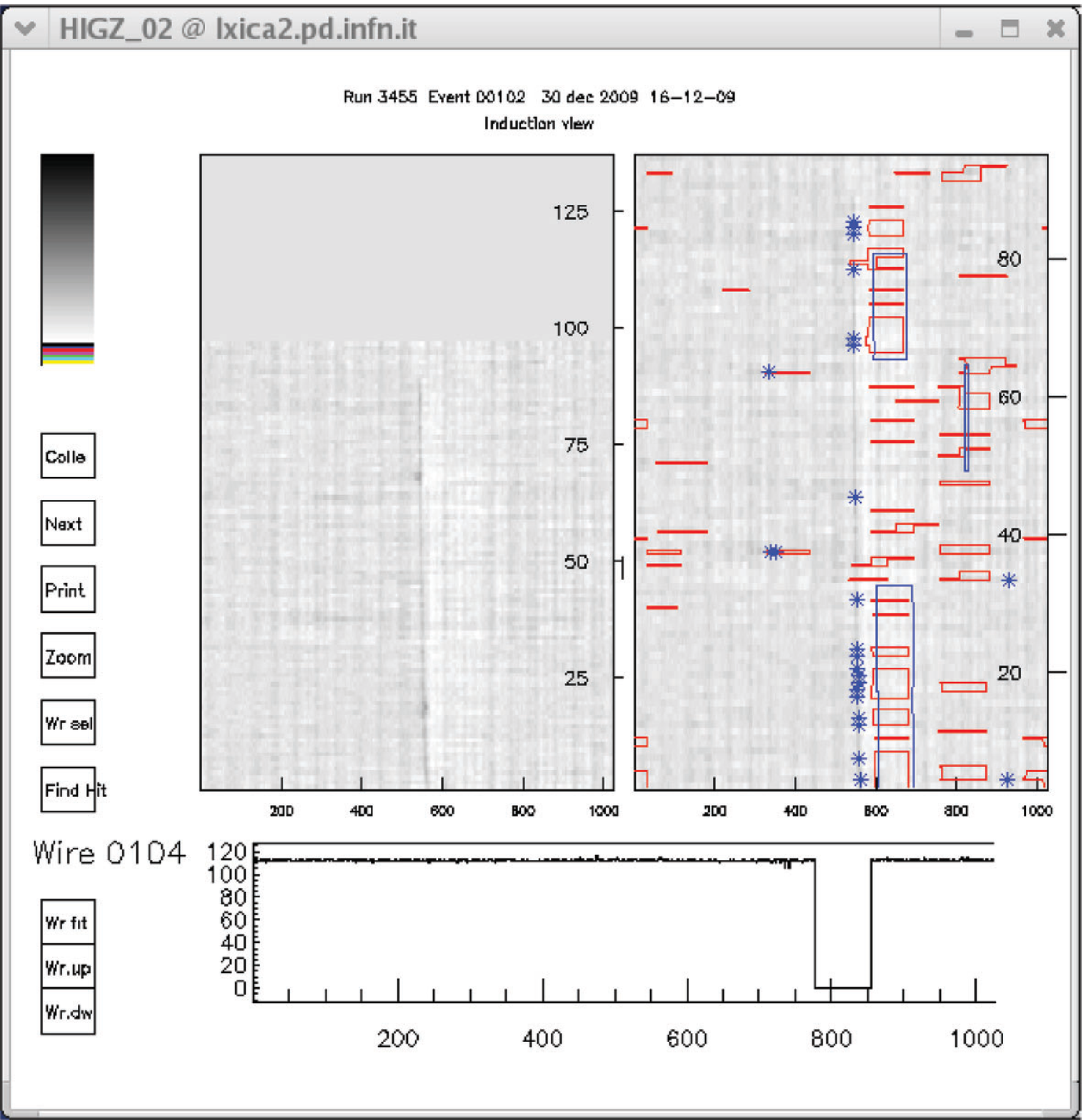}
\end{center}
\caption{Example of GTO signal inefficiency of the central Induction board for an vanishing 
muon track recorded with internal trigger. On the right it can be noticed that the number 
of hits identified by the DR-slw algorithm (red contour), even if larger than that one
 identified via software (blue stars), isn't enough to generate a GTO signal, which is 
 instead incorrectly generated by correlated noise (blue contour).}
\label{fig:event_display_peak_maj_Ind_bg}
\end{figure}

\noindent
The hit finding efficiency on the single wire was $\ge \ 97 \%$ up to 
threshold values $Q_{thr} = 6 \#$ ADC, like in Collection view, but  decreased more quickly 
with $Q_{thr}$.

\begin{figure}[tb]
 \begin{center}
 \includegraphics[width=0.7\textwidth]{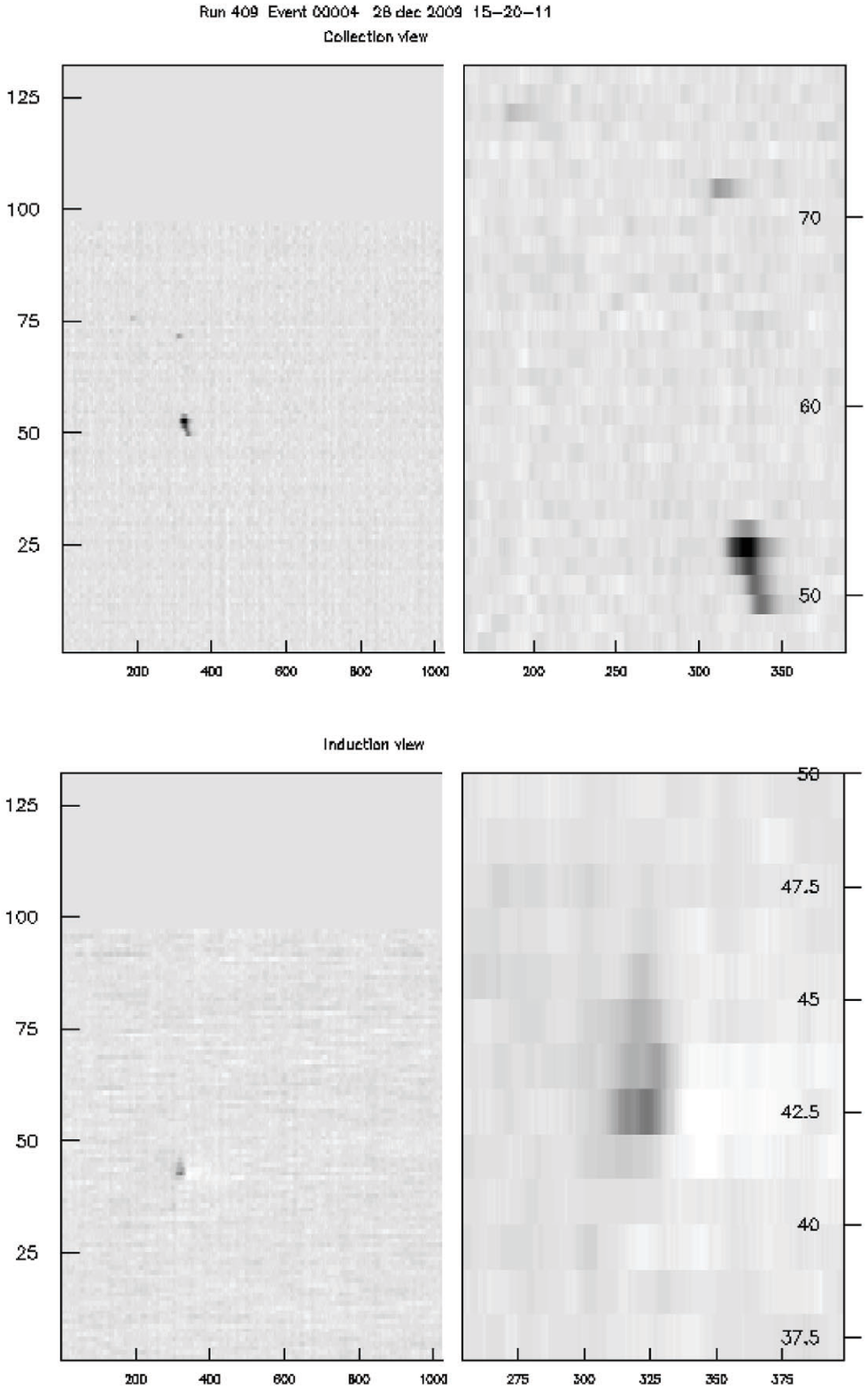}
\end{center}
 \caption{Isolated low energy event recorded in Icarino with the solar neutrino like internal 
 trigger, for $Q_{thr} = 6 \#$ ADC and $M = 4$, visualized in the Collection (top) and Induction 
 (bottom) views. On the right a zoomed view of the region interested by the event is shown.}
 \label{fig:event_nu_solar_1}
\end{figure}

  In order to avoid any  problem concerning unconfined muon tracks, the efficiency
  of the GTO signal from the central board in Induction view was measured with an upgraded internal trigger
  were by coincidence of the GTO signals from the the first and the third Induction boards was included.
 
  The obtained result, $\sim 98.0 \pm 0.3 \%$
  efficiency ($Q_{thr}= 5 \#$ ADC, $M = 6$ and stretching 
  $50\div 75 \ \mu$s) , is really satisfactory also because almost half of the small inefficiency 
  was due to vanishing tracks with few hits above threshold (fig. \ref{fig:event_display_peak_maj_Ind_bg}). 
  as recognized by a visual scan of the events.

\subsection{A low energy event trigger based on GTO signals}

\begin{figure}[tb]
 \begin{center}
  \includegraphics[width=0.7\textwidth]{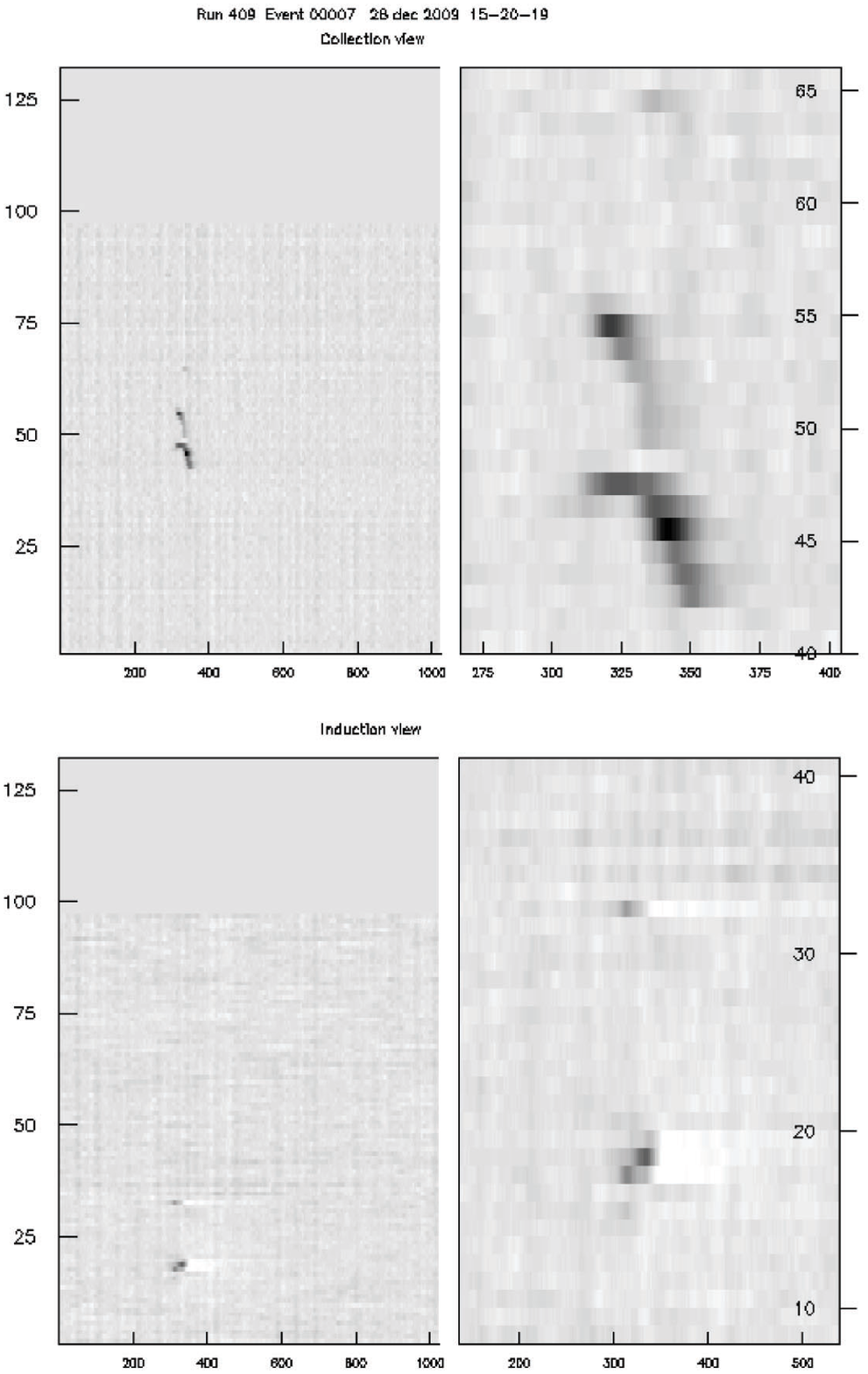}
\end{center}
 \caption{Isolated low energy event recorded in Icarino with the solar neutrino like internal
  trigger, for $Q_{thr} = 6 \#$ ADC and $M = 4$, visualized in the Collection (top) and Induction 
  (bottom) views. On the right a zoomed view of the region interested by the event is shown.}
 \label{fig:event_nu_solar_2}
\end{figure}

\begin{figure}[tb]
 \begin{center}
  \includegraphics[width=0.7\textwidth]{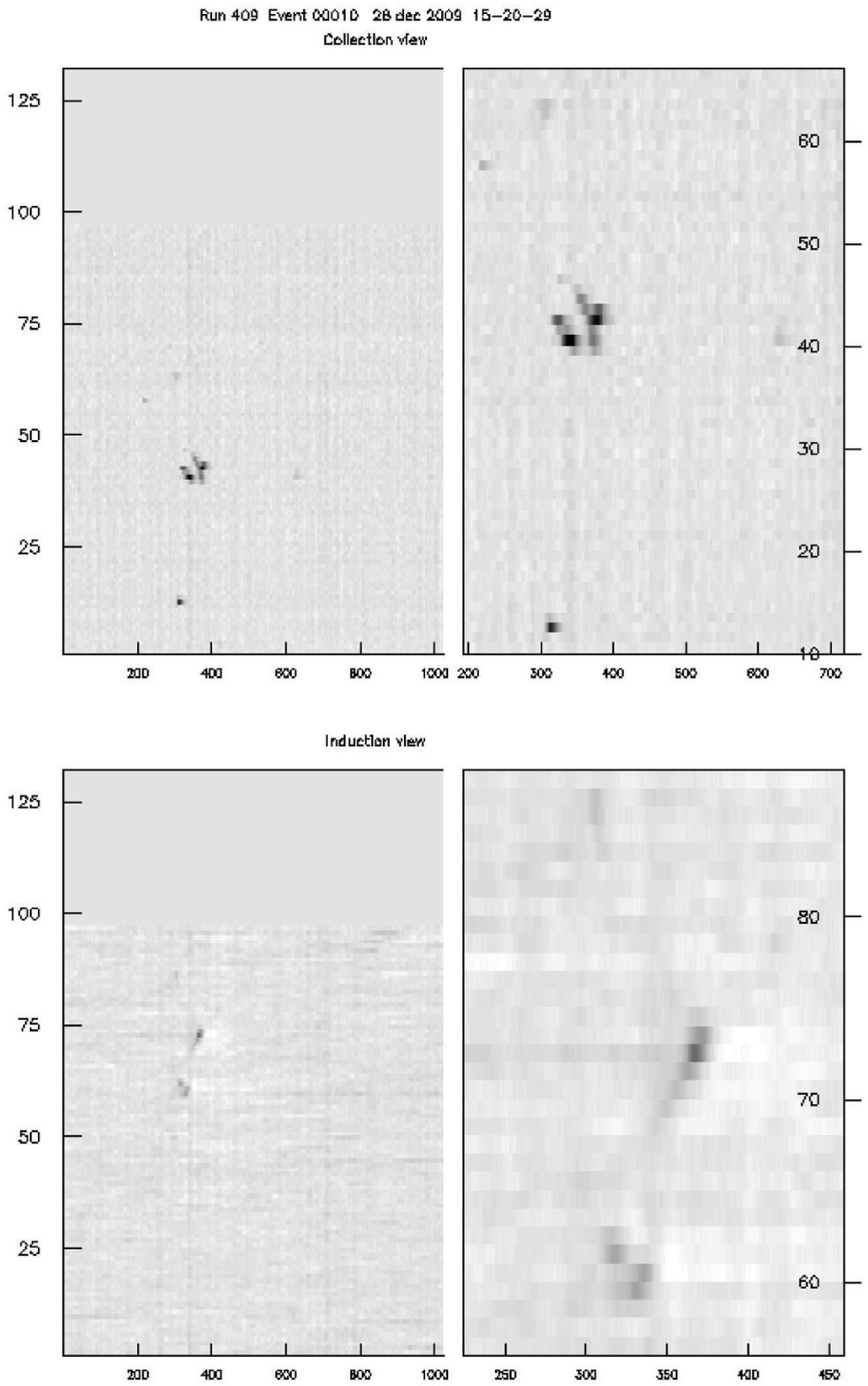}
\end{center}
 \caption{Isolated low energy event recorded in Icarino with the solar neutrino like internal 
 trigger, for $Q_{thr} = 5 \#$ ADC and $M = 3$, visualized in the Collection (top) and Induction 
 (bottom) views. On the right a zoomed view of the region interested by the event is shown.}
 \label{fig:event_nu_solar_3}
\end{figure}

\noindent 
A challenging item for huge mass detectors is the selection of localized low energy events, such as for 
example that ones involved in the search for solar or Supernovae neutrinos interactions.
Thus an interesting application of the internal trigger based on TPC wire signals was realized in 
Icarino vetoing the GTO signal from the central board with the coincidence of the GTO signals from the 
two lateral boards. The collected events, mainly $\gamma$ rays and neutrons 
(Figs. \ref{fig:event_nu_solar_1}, \ref{fig:event_nu_solar_2} and \ref{fig:event_nu_solar_3}), 
show on average a hit multiplicity per event in Collection(Induction)
 view $n_h = 7.3(7.9)$, for $Q_{thr} = 5 \#$ ADC and $M = 3$, and
  $n_h = 9.1(14.6)$, for $Q_{thr} = 6 \#$ ADC and $M = 4$, on an overall statistics of 4857 
  and 21483 events respectively.

\begin{figure}[tb]
 \begin{center}
   \includegraphics[width=0.49\textwidth]{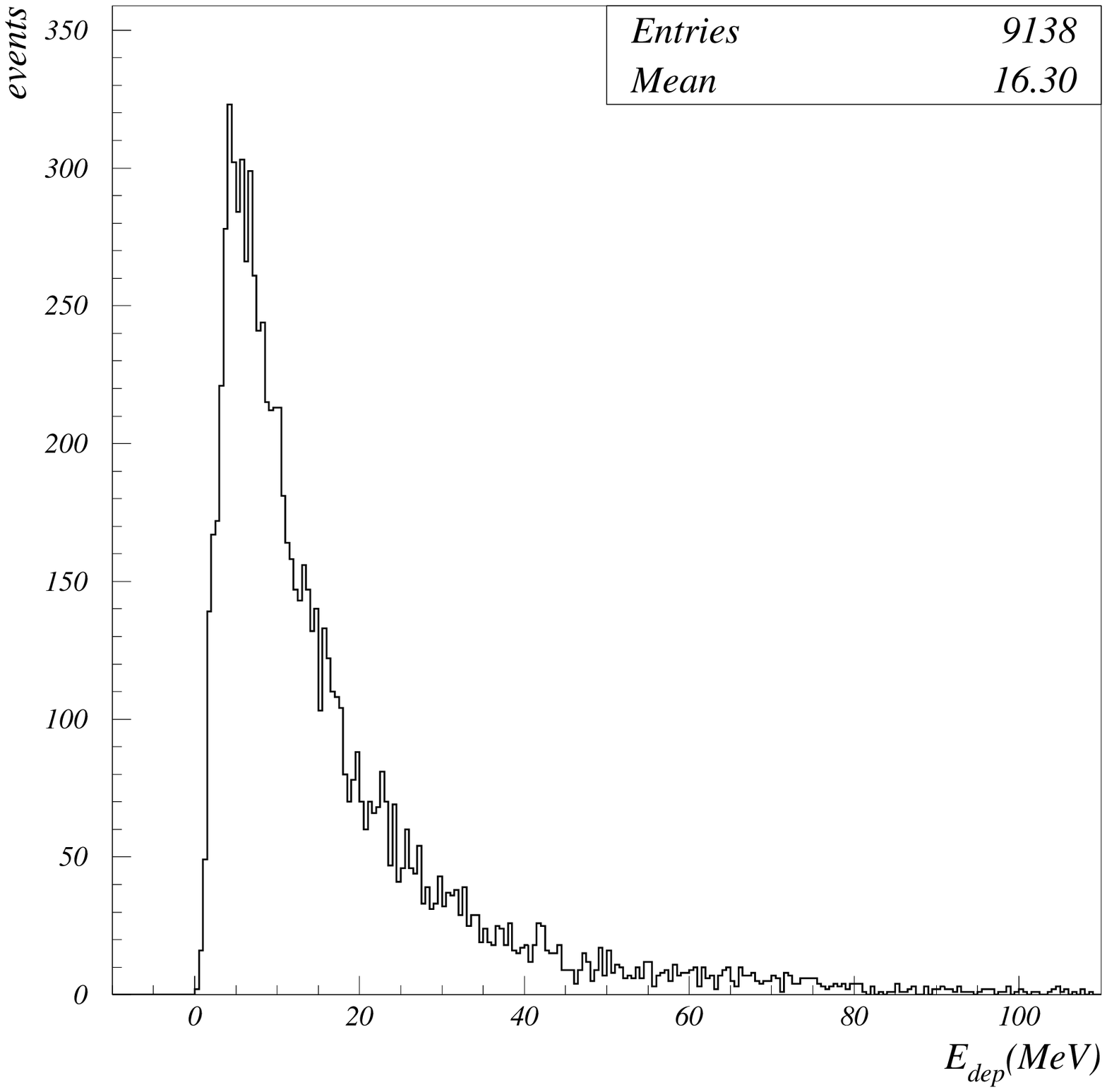}
   \includegraphics[width=0.49\textwidth]{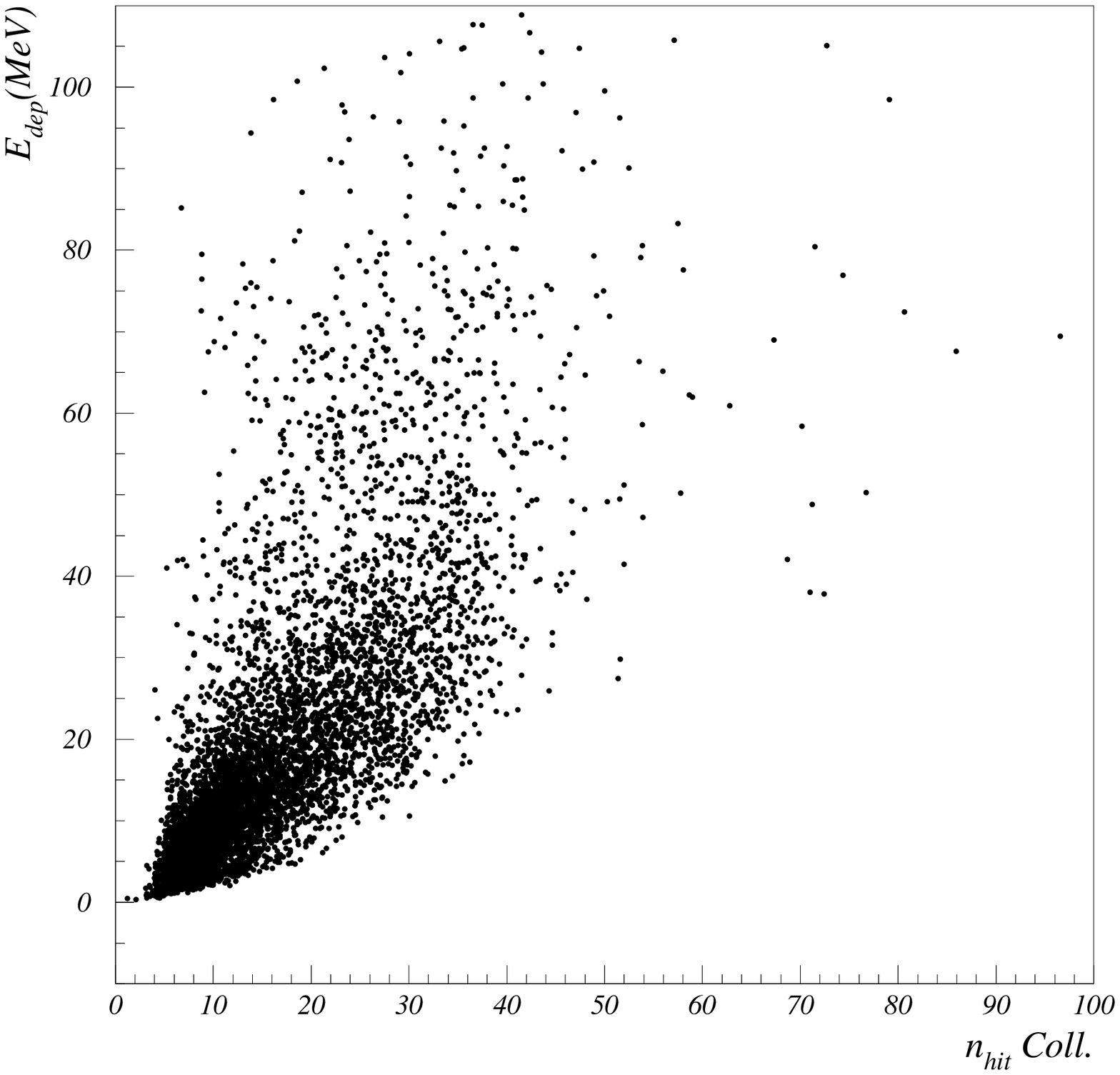}
   \includegraphics[width=0.49\textwidth]{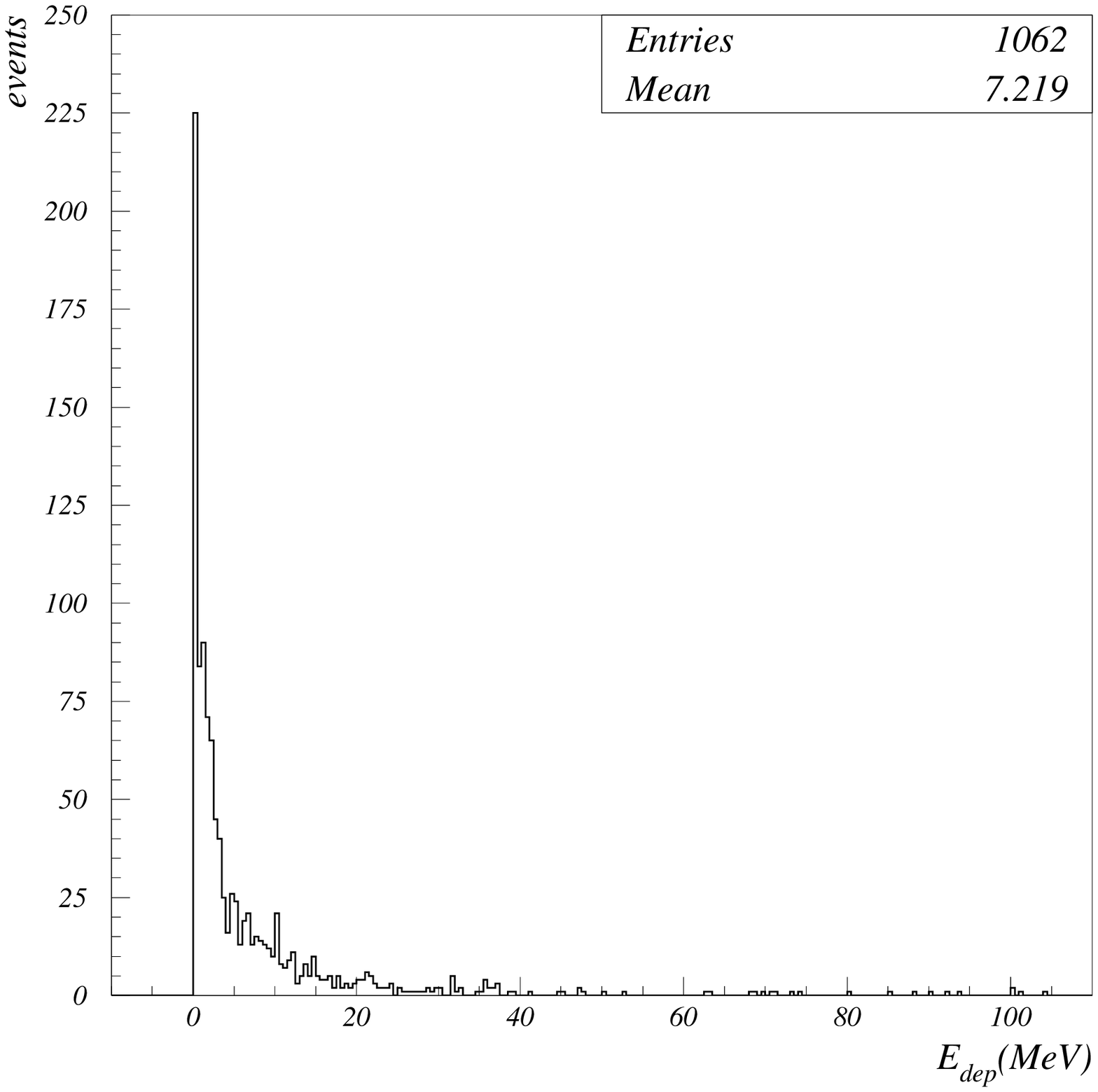}   
   \includegraphics[width=0.49\textwidth]{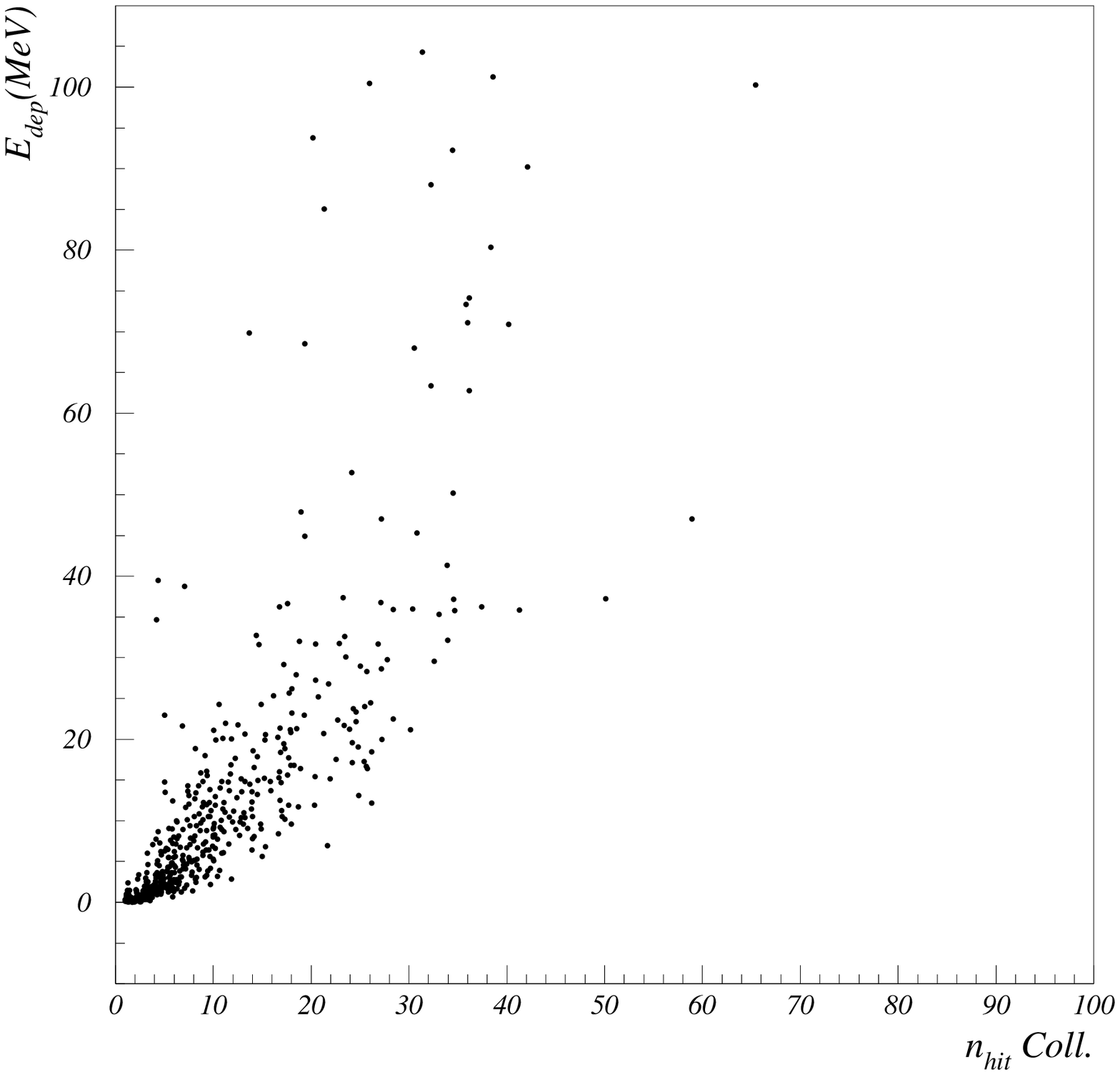}
\end{center}
 \caption{Deposited energy spectrum in Collection view $E_{dep}$ (left) and corresponding $n_{hits},\, E_{dep}$ 
 distribution of events recorded with the solar neutrino like internal trigger,
 for $Q_{thr} = 6 \#$ ADC 
 and M = 4 (top) and $Q_{thr} = 5 \#$ ADC and M = 3 (bottom).}
 \label{fig:Edep_nu_solar}
\end{figure}

\noindent
The deposited ionization energy $E_{dep}$ was measured event by event from the integral of the 
wire signal recorded in Collection view $A_i$, according to 
$E_{dep}=\Sigma_i A_i \frac{C}{e} \cdot \frac{E_{ion}}{R}$, where $e=1.6 \cdot 10^{-4}$ fC is the 
electron charge and $E_{ion}=23.6$ eV is the ionization energy in LAr. 
$C \ = \ (1.39 \pm 1 \%) \cdot 10^{-2}$ fC/(\# ADC $\cdot$ t-sample) is the on electronic chain calibration 
constant, extracted in a dedicated test-pulse run. Corrections to $E_{dep}$ from electron recombination 
was included (the attenuation due to Argon impurities is negligible because of the short drift path).
The resulting energy spectra of the events collected in the two runs extended up to 100 MeV 
(Fig. \ref{fig:Edep_nu_solar}), with an average value of 6.9 and 15.6 MeV respectively; the most 
energetic events referred  to inclined muon tracks with large $\delta$-rays traversing only the central 
region of the TPC.

\section{Conclusions}

The realization of a multikton mass LAr-TPC detector requires the extraction and the treatment of the corresponding huge quantity of data
from some $10^5$ channels. To increase the troughtput of the DAQ system an on-line lossless factor $\sim 3.9$  data compression based on the 
recording of the difference between the wire signal at two consecutive time sample was realized and tested with cosmic rays at the 
INFN LNL ICARINO test-facility.
  
A new algorithm based on a double rebinning - sliding window technique was studied and implemented on the read-out board for 
a full efficiency detection of a single hit signal on a TPC wire. In such a way a local trigger system 
was successfully implemented and tested with cosmic rays allowing triggering isolated low energy events down to 1 MeV visible energy 
using directly the wire signals. Moreover it's allow to define local ROIs (Region of Interest) avoiding a full acquisition 
of the detector when the event is localized.

\acknowledgments

The work described in  this paper has been carried out in the framework of the PRIN 2007 (Progetto di ricerca scientifica di Rilevante Interesse Nazionale) funded by the MIUR (Ministero dell'Istruzione, dell'Universita' e della Ricerca). The project, whose title was "Techniques for signal detection and treatment in large Liquid Argon apparatuses for new generation astro-particle physics experiments", involved the Universities and INFN section of L'Aquila, Padova, Pavia.

\end{document}